%% file: VHjet_akt.tex
\documentclass[a4paper,11pt]{article}
\usepackage{jheppub}
\usepackage[T1]{fontenc}
\usepackage{enumerate}
\usepackage{hyperref}
\usepackage{multirow}
\usepackage{lscape}
\usepackage{tabularx}
\usepackage[tableposition=top]{caption}

\newcommand{\mr}[1]{\mathrm{#1}}
\renewcommand{\d}{\mathrm{d}}
\newcommand{\as}{\alpha_s}
\newcommand{\asb}{\bar{\alpha}_s}

\def\R{{\scriptscriptstyle\mathrm{R}}}
\def\V{{\scriptscriptstyle\mathrm{V}}}
\def\MC{{\scriptscriptstyle\mathrm{MC}}}
\def\inn{\text{in}}
\def\out{\text{out}}

\def\CF{\mathrm{C_F}}
\def\CFsq{\mathrm{C_F^2}}

\def\CA{\mathrm{C_A}}
\def\CAsq{\mathrm{C_A^2}}
\def\CAcub{\mathrm{C_A^3}}
\def\CAfour{\mathrm{C_A^4}}

\def\cA{\mathcal{A}} \def\cAb{\mathcal{\bar{A}}}
\def\cB{\mathcal{B}} \def\cBb{\mathcal{\bar{B}}}
\def\cC{\mathcal{C}}

\def\cG{\mathcal{G}}
\def\cI{\mathcal{I}}
\def\cJ{\mathcal{J}}
\def\cK{\mathcal{K}}

 \def\cNb{\mathcal{\bar{N}}}
\def\cO{\mathcal{O}}
\def\cS{\mathcal{S}}

\def\oW{\overline{\mathcal{W}}}

\def\fA{\mathfrak{A}}

\def\s{\sigma}
\def\S{\Sigma}

\def\bA{\mathbf{A}}

\title{\boldmath Non-global logarithms up to four loops at finite-N$_c$ for V/H+jet processes at hadron colliders}

\author{Kamel Khelifa-Kerfa}

\affiliation{
Department of Physics, Faculty of Science and Technology\\
Universit\'{e} de Relizane, Relizane 48000, Algeria}
\affiliation{Laboratoire de Math\'{e}matique et Applications,\\
	Universit\'{e} Hassiba BenBouali de Chlef, Chlef 02000, Algeria}

\emailAdd{kamel.khelifakerfa@univ-relizane.dz}

\abstract{
We extend our previous work \cite{Khelifa-Kerfa:2015mma} on calculating non-global logarithms in $e^+ e^-$ annihilation to Higgs/vector boson production in association with a single hard jet at hadron colliders. We analytically compute non-global coefficients in the jet mass distribution up to four loops 
using the anti-k$_t$ jet algorithm. Our calculations are performed in the eikonal approximation, assuming strong energy ordering for the emitted gluons, thus capturing only the leading logarithms of the distribution. We compare our analytical results with the all-orders large-N$_c$ numerical solution. In general, the gross features of the non-global logarithm distribution observed in the $e^+ e^-$ case remain valid for the V/H+jet processes.
}

\keywords{QCD, Resummation, Jets, LHC, Higgs, Eikonal}

\arxivnumber{}

\begin{document}
\maketitle
\flushbottom

\section{Introduction}
\label{sec:intro}
   
In the endeavour of enhancing the possibility of disentangling new physics signals from those that are purely background, jet substructure techniques have played a central role, particularly at hadron colliders such as the current highest-energy CERN Large Hadron Collider (LHC). This is mainly due to the fact that at such colliders, final-state particles, including those resulting from the decay of Beyond Standard Model (BSM) heavy particles, are produced highly collimated and will most likely be reconstructed by jet algorithms into a single jet. Not only this, but jet substructure has also aided in the scrutiny of the Standard Model (SM) itself, especially in threshold phase space regions. Moreover, it has helped in initiating relatively new computational techniques in the field of high-energy physics, such as Machine Learning, as well as in providing a common ground for both theorists and experimentalists to enrich their interactions and discussions (see, for instance, Ref. \cite{Kogler:2018hem} for an experimental review and Ref. \cite{Larkoski:2017jix} for a theoretical review of jet substructure).

Amongst the widely studied jet substructures/shapes, both in $e^+ e^-$ and hadron collisions, is the invariant mass of a jet $m_j$. It is an infrared and collinear safe observable that is sensitive to soft and collinear emissions from both initial- and final-state partons, and thus important as a probe of various QCD aspects such as colour flow, hadronisation, underlying events, large logarithms, to name a few. It is part of a large class of observables, called {\it non-global} \cite{Dasgupta:2001sh, Dasgupta:2002bw}, that are sensitive to specific regions of phase space. These latter observables are notorious for giving rise to single logarithms\footnote{These are of the form $\alpha_s^n L^n$ where $n$ is an integer, $\alpha_s$ is the strong coupling constant, and $L$ is a large logarithm of the ratio of $m_j$ to another hard scale in the process, usually the transverse momentum of the jet $p_t$ or the centre-of-mass energy $Q$.} and beyond, in their perturbative series, known as {\it non-global logs} (NGLs), which are generally large and hence unavoidable for any precision calculation. They manifest the non-abelian nature of QCD and have not been fully understood despite the significant efforts by the jet substructure community (see, for example, \cite{Dasgupta:2002dc, Banfi:2002hw, Appleby:2002ke, Forshaw:2006fk, Banfi:2010pa, Khelifa-Kerfa:2011quw, Dasgupta:2012hg, Delenda:2012mm, Schwartz:2014wha} and references in \cite{Larkoski:2017jix}). Their state-of-the-art resummation is next-to-leading (NL) NGLs in the large-N$_c$ limit \cite{Banfi:2021owj, Banfi:2021xzn, Becher:2021urs, Becher:2023vrh} (N$_c$ is the number of quark colours) both for $e^+ e^-$ and hadronic colliders. 
Monte Carlo codes, such as \texttt{gnole} \cite{Banfi:2021owj, Banfi:2021xzn}, have been developed for the resummation of NGLs beyond leading-logarithm accuracy, addressing a wide range of observables. Furthermore, significant progress has been made towards constructing parton showers that achieve next-to-next-to-leading logarithmic (NNLL) accuracy, such as   \texttt{PanScales}, \cite{vanBeekveld:2023lsa, vanBeekveld:2023ivn, FerrarioRavasio:2023kyg, vanBeekveld:2024wws}. 
Meanwhile, there have been several ongoing efforts to resum NGLs beyond leading-colour accuracy (see, for instance, \cite{Hatta:2013iba, Hatta:2020wre, Hagiwara:2015bia, Platzer:2013fha,Forshaw:2019ver, AngelesMartinez:2018cfz, DeAngelis:2020rvq}) and for various jet algorithms \cite{Becher:2023znt}.

In the context of higher-order calculations, we computed in Ref. \cite{Khelifa-Kerfa:2015mma} NGL coefficients that are present in the hemisphere mass distribution in $e^+ e^-$ collisions fully up to 4-loops and partially through 5-loops. The calculations included full colour dependence, instead of the large-N$_c$ approximation that had been previously widely used. They were performed in the eikonal (soft) approximation and thus only guaranteed single logarithmic accuracy. We then computed NGLs up to 2-loops for hadron collisions in the anti-$k_t$ algorithm \cite{Cacciari:2008gp} in Ref. \cite{Dasgupta:2012hg} (Z+jet and dijet events), then in the $k_t$ \cite{Catani:1993hr, Ellis:1993tq} and Cambridge-Aachen (C-A) \cite{Dokshitzer:1997in} algorithms in Ref. \cite{Ziani:2021dxr} (Higgs/vector boson + jet events).

In the present paper, we generalise the work of \cite{Khelifa-Kerfa:2015mma} in a number of ways: first, we study the invariant mass of the leading-$p_t$ jet, instead of the hemisphere mass, and second, we consider scattering processes at hadron colliders, instead of $e^+ e^-$ annihilation. These generalisations make the current work much more subtle than that of \cite{Khelifa-Kerfa:2015mma}, due to the complexity of hadronic scattering environments (including initial-state radiation (ISR), parton distribution functions (PDFs), multiple emission dipoles, etc.), the presence of the jet radius parameter $R$ in all calculations, the various Born channels for a given process, to name a few. The specific hadronic processes that we shall be treating are the production of a Higgs boson $H$ or a vector boson ($Z, W^\pm$ or $\gamma$) in association with a single hard jet $j$. The state-of-the-art fixed-order QCD calculations for these processes are next-to-next-to-leading-order (NNLO) \cite{Gauld:2021ule, Boughezal:2015aha, Boughezal:2015ded, Gehrmann-DeRidder:2016cdi, Boughezal:2016dtm, Gehrmann-DeRidder:2015wbt, Campbell:2017dqk}. The resummation of NGLs at hadron colliders has only been performed numerically at large-N$_c$ in the anti-$k_t$ jet algorithm \cite{Dasgupta:2012hg} using the Monte Carlo (MC) program developed in \cite{Dasgupta:2001sh}.

We follow the procedure outlined in \cite{Khelifa-Kerfa:2015mma} by implementing the eikonal approximation and assuming strong energy-ordering of the transverse momenta of the radiated gluons; $p_t \gg k_{t1} \gg \cdots \gg k_{tn}$. This facilitates the computation of both real emission amplitudes and their corresponding virtual corrections. The general formalism as well as the detailed explicit formulae (up to 4-loops) for the eikonal amplitudes squared for the hadronic processes considered herein have been presented in our recent paper \cite{Khelifa-Kerfa:2020nlc}. The latter paper represents a generalisation of the previous $e^+ e^-$ calculations of eikonal amplitudes squared \cite{Delenda:2015tbo}. Furthermore, we apply the {\it measurement operator}, first introduced in \cite{Schwartz:2014wha}, to write the integrals of the NGL coefficients in a finite manner. The said integrals are then performed analytically whenever possible via a series expansion in the jet radius $R$, otherwise, we resort to multi-dimensional libraries, such as \texttt{Cuba} \cite{Hahn:2004fe, Hahn:2016ktb}, to compute them numerically. The final results for each Born channel are presented for 2-, 3-, and 4-loops. It is worth noting that all results shown herein are for the anti-$k_t$ jet algorithm. Similar calculations of NGLs in other jet algorithms will be presented elsewhere \cite{VHjet_kt_4loop}.

Furthermore, it has recently been demonstrated in \cite{Becher:2021zkk, Becher:2023mtx} that super-leading logarithms (SLLs), originally thought to only be present in \(2 \rightarrow 2\) processes \cite{Forshaw:2006fk, Forshaw:2008cq}, can arise in arbitrary \(2 \rightarrow l\) processes (where \(l = 0, 1, 2\)) and are therefore relevant to our V/H+jet process (a \(2 \rightarrow 1\) process). The authors also showed that SLLs first appear at the 4-loop order (\(\alpha_s^4\)) for \(2 \rightarrow 1\) processes. However, at this particular order in perturbation theory, NGLs and SLLs arise from distinct partonic configurations, and any potential interleaving between them, if it occurs, would begin at the 5-loop order (see, for instance, \cite{Forshaw:2006fk, Keates:2009dn, Becher:2023mtx}), which is beyond the scope of our current work.
Moreover, it has been shown that NGLs are generally more significant than SLLs \cite{Forshaw:2006fk}, and that SLLs are additionally numerically suppressed for the particular \(2 \rightarrow 0\) and \(2 \rightarrow 1\) processes \cite{Becher:2021zkk, Becher:2023mtx}. Therefore, we will not consider them herein and will leave their calculations for forthcoming papers at the 5- and 6-loop orders.

This paper is organised as follows. In sec. \ref{sec:Kinematics}, we present the details of the hadronic processes considered, including the various Born channels, recall the definition of the jet mass observable, the anti-$k_t$ jet algorithm, and the general formula of the jet mass distribution. After briefly reviewing the calculations at 1-loop, which have been performed previously (see for instance \cite{Dasgupta:2012hg} and \cite{Ziani:2021dxr}), in sec. \ref{sec:2-4loopCalcs}, we present the details of the calculations of NGL coefficients at 2-, 3-, and 4-loops. Moreover, comparisons of both the exponential and expansion forms of our NGL calculations to the all-orders large-N$_c$ results from \cite{Dasgupta:2012hg} and the MC program \cite{Dasgupta:2001sh} are carried out in sec. \ref{sec:Comparisons}. These will be used to hint at the significance of the finite-N$_c$ corrections as well as that of the missing higher-order terms. Finally, we conclude in sec. \ref{sec:Conclusion}.

\section{Definitions and kinematics}
\label{sec:Kinematics}

The problem that we treat in this work is the computation of NGLs at single logarithmic accuracy that appear in the distribution of the invariant mass of the leading hard jet produced in association with a Higgs or one of the vector bosons, $Z, W$ or $\gamma$, at hadron colliders. We shall adopt the notation used in our recent paper \cite{Ziani:2021dxr} which dealt with the same observable and for the same processes (but only for up to 2-loops). We note that for QCD calculations all bosons are considered as (colour-neutral) singlets and thus do not explicitly enter the NGLs calculations. Therefore, all aforementioned processes are, at the parton level, of the type of three-hard coloured QCD partons/legs considered in \cite{Khelifa-Kerfa:2020nlc}. They may only differ in the specific Born channels. We further note that all hard partons are considered massless.

\subsection{Hadronic processes} 

\begin{figure}[ht!]
\centering
\includegraphics[scale=0.8]{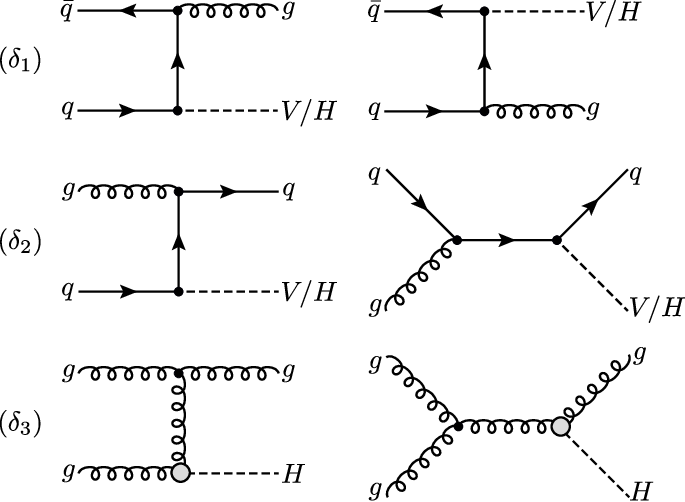}
\caption{Feynman diagrams for the three Born channels contributing to hadronic processes considered in the current work (figure from \cite{Ziani:2021dxr}).} \label{fig:FeynDiags}
\end{figure}

For the vector boson ($Z/W/\gamma$) + jet processes there are three Born channels that contribute to the scattering amplitude, namely: 
\begin{subequations}
\begin{align}\label{eq:BornChannels-1-2}
(\delta_1): q \bar{q} \to g + V, \qquad 
(\delta_2): q g \to q + V
\end{align}
and $\bar{q} g \to \bar{q}$, where $V$ refers to one of the vector (colour-neutral) bosons $Z, W$ or $\gamma$. The latter channel is in fact identical to ($\delta_1$) in terms of colour flow (which along with the Born cross section are the only distinguishing factors between all channels) and hence will not explicitly be considered further. The particular details of the production of $W^\pm$, which involves flavour changing, are too not relevant for our QCD NGLs calculations. Furthermore, the Higgs + jet production has in addition to the above mentioned three channels a fourth one, namely the all-gluons channel 
\begin{align}\label{eq:BornChannels-3}
(\delta_3): g g  \to g + H.
\end{align}
\end{subequations}
Fig. \ref{fig:FeynDiags} shows the Feynman diagrams corresponding to the above three Born channels. 

\subsection{Observable definition}

At the parton level we schematically represent the scattering process for the various Born channels mentioned above in association with the emission of $n$ soft gluons $g_i$ as  
\begin{align}\label{eq:PartonProcess+n_SoftGluons}
 a(p_a) + b(p_b) \to j(p_j) + X + g_1(k_1) + \cdots + g_n(k_n), 
\end{align}
where $a$ and $b$ label the incoming hard partons, $j$ the outgoing hard parton initiating the leading hard jet, $X$ the Higgs or $V$ bosons, and $p_i$ and $k_\ell$ the corresponding four-momenta of the $i^{th}$ hard parton ($a, b$ or $j$) and the $\ell^{th}$ soft gluon, respectively. The explicit expressions of the latter momenta are given by:
\begin{subequations}
\begin{align}
p_a &= x_a \frac{\sqrt{s}}{2} (1,0,0,1),  \\
p_b &= x_b \frac{\sqrt{s}}{2} (1,0,0,-1),  \\
p_j &= p_t (\cosh y, \cos\varphi, \sin\varphi, \sinh y), \\
k_i &= k_{ti} (\cosh \eta_i, \cos\phi_i, \sin\phi_i, \sinh \eta_i), 
\end{align} 
\end{subequations}
where $p_t, y$ and $\varphi$ are the transverse momentum, rapidity and azimuthal angle of outgoing final-state hard jet, and $k_{ti}, \eta_i$ and $\phi_i$ are the transverse momentum, rapidity and azimuthal angle of the $i^{th}$ soft emission, measured with respect to the beam axis. The collision centre-of-mass energy is $\sqrt{s}$ and $x_a, x_b$ are the momentum fractions carried by the incoming beam partons $a$ and $b$, respectively. Worth mentioning is that the effect of recoil of hard partons against soft emissions is beyond single logarithmic accuracy (see for instance Ref. \cite{Banfi:2004yd}) and will thus be neglected throughout.  

The (squared) invariant mass of the leading hard jet normalised to the its transverse momentum $p_t$ is given by the sum of the momenta of the hard outgoing massless parton $j$ and the final-state soft emissions $k_i$ that end up inside the jet after applying the jet algorithm. That is:  
\begin{align}\label{eq:JetMassSq_Normalised_Defn}
 \varrho &= \frac{m_j^2}{p_t^2} =  \frac{1}{p_t^2} \left(p_j + \sum_{i \in j} k_{i} \right)^2 = \sum_{i \in j} \varrho_i + \mathcal{O} \left(\frac{k_{t}^2}{p_t^2}\right), 
 \notag \\
 \varrho_i &= \frac{2 (p_j \cdot k_{i})}{p_t^2} = 2 \xi_i \left[\cosh(\eta_i - y) - \cos(\phi_i - \varphi)\right], 
\end{align}
where $\xi_i = k_{ti}/p_t$ and in the soft limit one neglects terms that are proportional to $\xi^2$ (or equivalently $k_t^2$). 

The detailed recipe of the three well-known jet algorithms is well explained, for example, in \cite{Ziani:2021dxr}. The anti-k$_t$ jet algorithm works, in the strong-energy ordering regime, in a simple manner. In the $(\eta, \phi)$ plane, one draws a circle of radius $R$ around the hard outgoing parton $j$. Any soft emission $k_i$ that resides inside this circle will be recombined with the latter jet, otherwise it will not. The momentum of the resultant jet will be the sum of the momenta of the constituent partons. Soft gluons that are not clustered as part of the said hard jet may be clustered with other gluons, if they are within a radius of $R$ from each other, otherwise they will be considered as separate final state jets. Mathematically, a soft gluon $k_i$ is considered inside the jet if the following condition is satisfied 
\begin{align}\label{eq:AntiKt_Condition}
   (y-\eta_i)^2 + (\varphi - \phi_i)^2 < R^2. 
\end{align}

\begin{figure}[ht]
	\centering
	\includegraphics[scale=0.6]{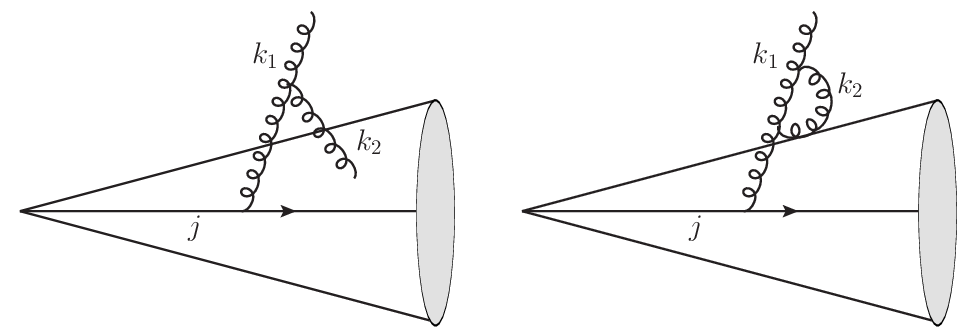}
	\caption{A schematic diagram showing how NGLs arise at 2-loops.} \label{fig:NGLs_2loop}
\end{figure}

As mentioned above, \eqref{eq:JetMassSq_Normalised_Defn}, only partons that are clustered with the hard jet contribute to its mass. The emergence of NGLs for the jet mass observable has been extensively discussed in prior literature (see, for example, \cite{Banfi:2010pa, Khelifa-Kerfa:2011quw, Dasgupta:2012hg, Khelifa-Kerfa:2015mma}). Essentially, the phenomenon begins with the radiation of two soft gluons, $k_1$ and $k_2$, from one of the Born configurations, Eqs. \eqref{eq:BornChannels-1-2} and \eqref{eq:BornChannels-3}. The first, harder gluon $k_1$, emitted outside the jet, subsequently emits the second, softer real gluon $k_2$ into the jet vicinity (see Fig. \ref{fig:NGLs_2loop}). This gluon $k_2$ then contributes to the mass of the jet. However, if gluon $k_2$ is virtual, it cannot be clustered with the jet and thus does not contribute to its mass. This scenario results in an incomplete cancellation between real and virtual contributions to the jet mass observable, leaving large logarithmic terms in the jet mass.

The details of the various configurations that give rise to NGLs at 2-, 3- and 4-loops have been thoroughly discussed in our previous work \cite{Khelifa-Kerfa:2015mma} and shall not be repeated herein. The interested reader is referred to the latter reference for more information. The expressions of the integrals of the NGLs coefficients at a given order in the perturbation expansion of the jet mass will be identical to those presented in the said reference for the same order (more details later).

A parametrisation in terms of the polar variables $(r, \theta)$ that will prove useful in performing NGLs integrals is given by 
\begin{align}\label{eq:PolarVariables_Param}
 \eta_i - y = R r_i \cos\theta_i, \qquad 
 \phi_i - \varphi = R r_i \sin\theta_i,
\end{align}
where $r_i >0$, $2\pi > \theta_i>0$ and $\d\eta_i \d\phi_i = R^2\,r_i \d r_i \d\theta_i$. 
The anti-k$_t$ clustering condition \eqref{eq:AntiKt_Condition} reduces, in the latter parametrisation, to the simple form: 
\begin{align}\label{eq:AntiKt_Condition_Polar}
 k_i \in j \qquad \Rightarrow \qquad  r_i < 1.
\end{align}   
Moreover, the normalised jet mass \eqref{eq:JetMassSq_Normalised_Defn} may be expanded as a series in $R$. That is: 
\begin{align}
  \varrho_i =  \xi_i \left[R^2 r_i^2 + \frac{1}{12} R^4 r_i^4 \cos(2\theta_i) + \frac{1}{2880} \left(5 + 3\cos(4\theta_i)  \right) R^6 r_i^6 +\cdots \right]. 
\end{align}
For the accuracy level of our calculations, which is single log, it suffices to only keep the first term in the above expansion. Hence, in the remaining of the paper, we shall assume, unless stated otherwise, that  
\begin{align}\label{eq:JetMass_Normalised_Approx}
\varrho_i \approx  \xi_i R^2 r_i^2. 
\end{align}

\subsection{Jet mass distribution}

The differential cross-section for the distribution of the normalised invariant jet mass may be written, for a specific channel $\delta$, as 
\begin{align}\label{eq:Xsec_Differential}
 \frac{\d \S_\delta (\rho)}{\d \cB_\delta} = \int_0^\rho \frac{\d^2 \s_{\delta}}{\d \cB_\delta \d \varrho} \d \varrho,
\end{align}
where $\cB_\delta$ denotes the Born configuration with a differential element given by $\d\cB_\delta = \d x_a\, \d x_b\, f_a(x_a, \mu_\mr{F}^2)\, f_b(x_b, \mu_\mr{F}^2)$, where $f_i$ is the PDF of the $i^{th}$ incoming parton and $\mu_\mr{F}$ is the factorisation scale (see appendix A of Ref. \cite{Ziani:2021dxr} for full details). 
In the small mass region, $\rho \ll 1$, where large logarithms dominate the distribution, the differential cross-section \eqref{eq:Xsec_Differential} may be cast in the form 
\begin{align}\label{Xsec_Differential_Form2}
  \frac{\d \S_\delta (\rho)}{\d \cB_\delta} = \frac{\d\s_{0,\delta}}{\d \cB_\delta}\, f_{\cB, \delta}(\rho)\, \left[1 + \mathcal{O}(\as)\right], 
\end{align}
where $\d\s_{0, \delta}/\d\cB_\delta$ is the partonic differential cross-section for the Born channel $\delta$ (discussed in depth in appendix A of \cite{Ziani:2021dxr}). The function $f_{\cB, \delta}(\rho)$ resums all various large logarithms appearing in the $\rho$-distribution. It may be written in the factorised form
\begin{align}\label{eq:ResumFF_FactorisedForm}
  f_{\cB, \delta}(\rho) = f_{\cB, \delta}^{\mr{global}}(\rho)\, \cS_{\delta}(\rho), 
\end{align}
where $f_{\cB, \delta}^{\mr{global}}(\rho)$ represents the well-known Sudakov form factor that results from the exponentiation of the single-gluon emission. It accounts for soft- and hard-collinear radiations off the outgoing jet as well as soft wide-angle primary (non-correlated) radiations from all hard partons (incoming and outgoing). It may be written, up to NLL, as \cite{Dasgupta:2012hg, Ziani:2021dxr}: 
\begin{align}\label{ResumFF_Global}
 f_{\cB, \delta}^{\mr{global}}(\rho) = \frac{ e^{- \mathcal{R}_\delta(\rho) - \gamma_E \mathcal{R}'_\delta(\rho)} }{\Gamma\left[1 + \mathcal{R}'_\delta(\rho) \right] },
\end{align}
where $\gamma_E \approx 0.577$ and the various terms have been defined and fully computed in \cite{Dasgupta:2012hg} with their explicit expressions presented in the appendix C of the said reference and appendix B of \cite{Ziani:2021dxr}.  

In the current paper we are interested in the term $\cS_\delta(\rho)$ which represents the resummation of leading (single) NGLs. Its fixed-order series expansion may be written as
\begin{align}\label{eq:NGLs_Resummed}
 \cS_{\delta}(\rho) = 1 + \cS_{2, \delta}(\rho) + \cS_{3, \delta}(\rho) + \cS_{4, \delta}(\rho) + \cdots.  
\end{align} 
In what follows below we will be carrying out detailed calculations of the first three NGLs contributions above. i.e., at 2-, 3- and 4-loops.

\section{NGLs calculations}
\label{sec:2-4loopCalcs}

\subsection{NGLs at 2-loops}

As we mentioned above, the integrals of the NGLs coefficients for our hard processes assume the same forms as those encountered in $e^+ e^-$ calculations \cite{Khelifa-Kerfa:2015mma}. In other words, the 2-loops NGLs integral reads 
\begin{align}\label{eq:NGLs_2loop_Integral}
 \cS_{2, \delta}(\rho) = -\int_{\xi_1 > \xi_2} \d \Pi_{12}\, \Xi^{\mr{anti-k}_t}(k_1, k_2)\, \oW_{12, \delta}^{\R\R}, 
\end{align}  
where $\Xi^{\mr{anti-k}_t}(k_1, k_2)$ is the constraint resulting from the application of the anti-k$_t$ algorithm. It reads 
\begin{align}\label{eq:2loop_akt_Constraint}
  \Xi^{\mr{anti-k}_t}(k_1, k_2) &= \Theta\left[(\eta_1 - y)^2 + (\phi_1 - \varphi)^2 - R^2\right] \Theta\left[R^2 - (\eta_2 - y)^2 - (\phi_2 - \varphi)^2\right], \notag
  \\
  & \equiv \Theta^\out_1 \Theta^\inn_2. 
\end{align}
The 2-loops phase space factor $\d\Pi_{12} = \d\Phi_1 \d\Phi_2 \Theta(\varrho_1 - \rho) \Theta(\varrho_2 - \rho)$ where the 1-loop phase space factor  is given by
\begin{align}\label{eq:1loop_PhaseSpaceFactor}
 \d\Phi_i = \asb \, \frac{\d \xi_i}{\xi_i}\, \d\eta_i \, \frac{\d\phi_i}{2\pi}, 
\end{align}
with $\asb = \as/\pi$. The ({\it irreducible} part) of the eikonal amplitude squared $\oW_{12, \delta}^{\R\R}$ appearing in \eqref{eq:NGLs_2loop_Integral} is given in \cite{Khelifa-Kerfa:2020nlc}: 
\begin{align}\label{eq:2loop_EikAmp_Irred}
 \oW_{12, \delta}^{\R\R} = \CA \, \sum_{(ij) \in \Delta_\delta} \cC_{ij} \, \cA_{ij}^{12}, 
\end{align}
where the sum is over all possible dipoles formed by the partons in the Born channel $\delta$. That is,  $\Delta_\delta = \{(aj), (bj), (ab) \}$. The dipole colour factors $\cC_{ij}$ have been discussed in Refs. \cite{Khelifa-Kerfa:2020nlc} and \cite{Delenda:2015tbo}. They read: 
\begin{align}\label{eq:1loop_ColorFactors}
 \cC_{q\bar{q}} = \cC_{qq} = 2\CF-\CA, \qquad \cC_{qg} = \cC_{gg} = \CA, 
\end{align}
where $\CF = (N_c^2-1)/2N_c$ and $\CA = N_c$ are the colour Casimir scalars for (anti)quarks and gluons, respectively. The 2-loops antenna function $\cA_{ij}^{12}$ is defined, in a general form, as
\begin{align}\label{eq:2loop_AntennaFun}
  \cA_{\alpha \beta}^{ij} = w_{\alpha \beta}^i (w_{\alpha i}^j + w_{i \beta}^j - w_{\alpha \beta}^j),
\end{align}
where $w_{\alpha \beta}^i$ is the 1-loop dipole antenna function defined by \footnote{Note that compared to the definitions of the phase space factor and the 1-loop antenna function in \cite{Khelifa-Kerfa:2020nlc} a factor of $1/2$ has been moved from the former to the latter in the present paper.}
\begin{align}\label{eq:1loop_AntennaFun}
 w_{\alpha \beta}^i = \frac{k_{ti}^2}{2}\, \frac{(p_\alpha \cdot p_\beta)}{(p_\alpha \cdot k_i)(k_i \cdot p_\beta)}. 
\end{align}
Notice that both the 1-, and 2-loops antenna functions are purely angular functions. That is, they depend neither on the colour flow nor on the four momenta. 

Substituting Eqs. \eqref{eq:2loop_akt_Constraint}, \eqref{eq:1loop_PhaseSpaceFactor} and \eqref{eq:2loop_EikAmp_Irred} back into the integral of the 2-loops NGLs coefficient \eqref{eq:NGLs_2loop_Integral} we have 
\begin{align}\label{eq:NGLs_2loop_Integral2}
 \cS_{2,\delta}(\rho) &= - \CA \sum_{(ij) \in \Delta_\delta} \cC_{ij} \, \asb^2\, \int_{\xi_1 > \xi_2} \frac{\d\xi_1}{\xi_1} \frac{\d\xi_2}{\xi_2}\, \d\eta_1 \d\eta_2 \,  \frac{\d\phi_1}{2\pi} \frac{\d\phi_2}{2\pi}\, \Theta_1^\out \Theta_2^\inn \times 
 \notag\\
 &\times \Theta(\varrho_1 - \rho) \Theta(\varrho_2 - \rho) \, \cA_{ij}^{12}. 
\end{align}
In order to perform the above integration we use the  change of variables \eqref{eq:PolarVariables_Param} and the accompanying approximations \eqref{eq:AntiKt_Condition_Polar} and \eqref{eq:JetMass_Normalised_Approx}, to write 
\begin{align}\label{eq:NGLs_2loop_Integral3}
 \cS_{2,\delta}(\rho) &= - \CA \sum_{(ij) \in \Delta_\delta} \cC_{ij} \, \asb^2\, R^4 \int_{\xi_1 > \xi_2} \frac{\d\xi_1}{\xi_1} \frac{\d\xi_2}{\xi_2}\, r_1\d r_1\, r_2 \d r_2 \,  \frac{\d\theta_1}{2\pi} \frac{\d\theta_2}{2\pi}\, \Theta(r_1 -1) \Theta(1 - r_2) \times 
\notag\\
&\times \Theta(\xi_1 r_1^2 R^2 - \rho) \Theta(\xi_2 R^2 r_2^2 - \rho) \, \cA_{ij}^{12}. 
\end{align}
Up to single log accuracy, i.e., keeping only the leading NGLs, the energy $\xi$'s integrals factorise out to give $L^2/2!$ where $L = \ln(R^2/\rho)$. We may then write the 2-loops NGLs contribution as:
\begin{align}\label{eq:NGLs_2loop_FinalForm}
 \cS_{2,\delta}(\rho) = -\frac{1}{2!} \, \asb^2\, L^2\, \cG_{2,\delta}(R), 
\end{align}
where the NGLs coefficient
\begin{align}\label{eq:NGLs_2loop_G2}
\cG_{2,\delta}(R) &= \CA \sum_{(ij) \in \Delta_\delta} \cC_{ij} \, R^4 \int_1^{\frac{\pi}{R |\sin\theta_1|}} r_1\d r_1 \, \int_0^{2\pi} \frac{\d\theta_1}{2\pi}\,\int_0^1 r_2 \d r_2 \,   \int_0^{2\pi} \frac{\d\theta_2}{2\pi}\,  \cA_{ij}^{12}, 
\notag\\
&=  \CA \sum_{(ij) \in \Delta_\delta} \cC_{ij}\, \cI_{ij}(R).  
\end{align}
Few points to mention regarding the above integral. First, up to single log accuracy the lower limit on $r_2$'s integral, which is $\sqrt{\rho/R^2}$, has been set to $0$ since the 2-loops antenna function $\cA_{ij}^{12}$ is finite. Second, the upper limit on the $r_1$ integral comes from the fact that $(\phi_1 - \varphi) \in [-\pi, \pi]$ and from Eq. \eqref{eq:PolarVariables_Param} it follows that $\pi/(R \sin\theta_1) > r_1 >-\pi/(R \sin\theta_1)$. Since $r_1 > 1$ (from anti-k$_t$ clustering condition \eqref{eq:AntiKt_Condition_Polar}) and $\sin\theta_1$ changes its sign over the range $[0,2\pi]$ one finds the upper limit shown in Eq. \eqref{eq:NGLs_2loop_G2}. 

\begin{figure}[h!]
	\centering
	\includegraphics[scale=0.59]{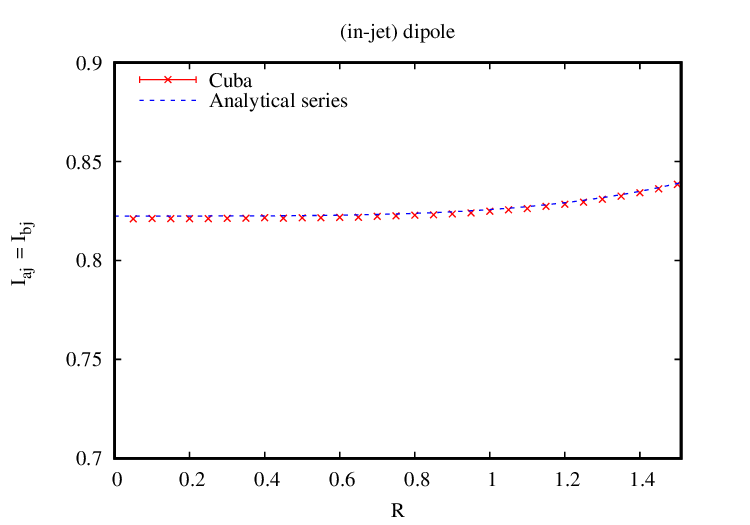}
    \includegraphics[scale=0.59]{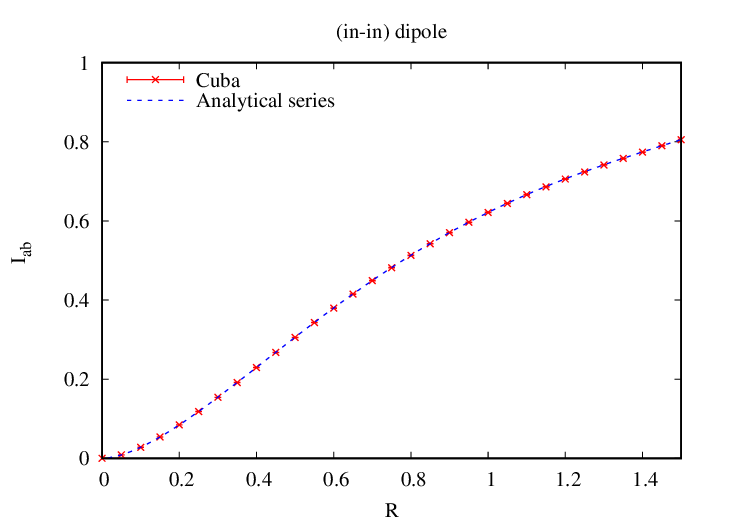}	
	\caption{Comparisons between analytical and numerical results for the NGLs coefficients 2-loops.} \label{fig:NGLs_2loop_Iaj-ab}
\end{figure}

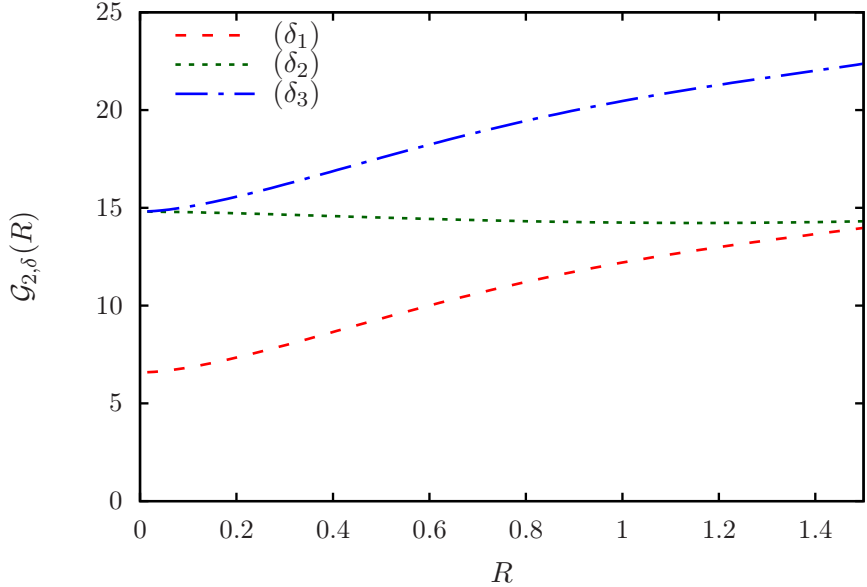
\begin{figure}[h!]
	\centering
	\input{Figs/G2.tex}
	\caption{NGLs coefficients at 2-loops for all of the three Born channels.} \label{fig:NGLs_2loop_G2}
\end{figure}

The procedure that we shall follow to do the angular integrals $\cI_{ij}(R)$ of Eq. \eqref{eq:NGLs_2loop_G2}, as well as the corresponding integrals at 3- and 4-loops, is as follows: 
\begin{enumerate}
\item Compute the integral for each dipole ($ij$) separately. 
 \item Substitute the parametrisation \eqref{eq:PolarVariables_Param} into the expression of the angular integrand (the 2-loops antenna function $\cA_{ij}^{12}$ in \eqref{eq:NGLs_2loop_G2}) and expand it as a series in $R$. 
 \item Whenever the upper limit is of the form $\pi/(R |\sin\theta|)$ then split the integration range into two regions: (a) $1 <r < \pi/R$, and (b) $\pi/R < r < \pi/(R |\sin\theta|) $. 
 \item Sum up the contributions from all regions. 
\end{enumerate}
We note that all of our analytical series results for all integrals have been verified numerically using, as stated above, \texttt{Cuba} library. We find the following expressions for the NGLs coefficients $\cI_{ij}$ for the (incoming-jet) dipole, labelled as $(aj)$ or $(bj)$,  and (incoming-incoming) dipole, labelled as $(ab)$, dipoles:
\begin{subequations}
\begin{align}
 \cI_{aj} = \cI_{bj} &= \frac{\zeta_2}{2} + 0.003\,R^4 + \cO(R^8), 
 \label{eq:NGLs_2loop_Iaj}\\
 \cI_{ab} &= \frac{1}{2} R^2 \left(1 - 2 \ln R\right) + \frac{1}{8} R^4 - 0.003 R^6 + \cO(R^8).
 \label{eq:NGLs_2loop_Iab}
\end{align}
\end{subequations}
Fig. \ref{fig:NGLs_2loop_Iaj-ab} shows comparisons between the analytical series above,  \eqref{eq:NGLs_2loop_Iaj} and \eqref{eq:NGLs_2loop_Iab}, and the pure numerical integration results. Our findings agree with those reported previously \cite{Dasgupta:2012hg, Ziani:2021dxr}. 
Substituting Eqs. \eqref{eq:NGLs_2loop_Iaj} and \eqref{eq:NGLs_2loop_Iab} back into the expression of $\cG_{2,\delta}$ \eqref{eq:NGLs_2loop_G2} and simplifying we obtain the following formulae for each of the three channels: 
\begin{subequations}
	\label{eq:NGLs_2loop_G2_FinalChannel}
\begin{align}
 \cG_{2,\delta_1}(R) &= \CF \CA \left[1.645 + 0.007 R^4\right] 
                                        + \CAsq \left[ 0.5 R^2 - R^2 \ln(R) + 0.125 R^4 - 0.003 R^6\right]  + \notag \\ &+ \cO(R^8),
 \\
 \cG_{2,\delta_2}(R) &= \CF \CA \left[R^2 -2 R^2 \ln(R)+0.25 R^4-0.007 R^6\right] +\notag\\
                                    &  + \CAsq \left[1.645 - 0.5 R^2 +R^2 \ln(R) - 0.118 R^4 + 0.003 R^6 \right] + \cO(R^8),
 \\
 \cG_{2,\delta_3}(R) &= \CAsq \left[ 1.645 + 0.5 R^2 -  R^2 \ln(R) + 0.132 R^4 - 0.003 R^6 \right] + \cO(R^8).
\end{align}
\end{subequations}
In Fig. \ref{fig:NGLs_2loop_G2} we show the NGLs coefficients \eqref{eq:NGLs_2loop_G2_FinalChannel} for all three channels. 
Clearly the contribution of channel $(\delta_3)$ is the largest due to the colour factor. Note that in the limit $R \to 0$ one finds for the above NGLs coefficients: 
\begin{align}
\lim_{R\to 0} \cG_{2,\delta_1}(R) = \CF\CA\, \zeta_2, \qquad 
\lim_{R\to 0} \cG_{2,\delta_2}(R) = \CAsq\, \zeta_2, \qquad 
\lim_{R\to 0} \cG_{2,\delta_3}(R) = \CAsq\, \zeta_2,
\end{align}
where $\zeta_2 = 1.645$. i.e., one recovers the results found in $e^+ e^-$ hemisphere mass distribution \cite{Dasgupta:2001sh, Khelifa-Kerfa:2015mma}. This is due to the fact that NGLs arise predominantly at the boundary of the phase space region of interest. In other words, as far as the jet region, the phase space region relevant to our jet mass observable, has a boundary (or edge) even if it is vanishingly small then NGLs effects will be present.  

The calculations presented thus far, for NGLs at 2-loops, are not entirely new and have already been presented in the literature. It is the 3- and 4-loops calculations, discussed in the next sections, that are new and presented herein for the first time in the literature.

\subsection{NGLs at 3-loops}

\begin{figure}[h!]
	\centering
    \input{Figs/G3.tex}
	\caption{NGLs coefficients at 3-loops for all of the three Born channels.} \label{fig:3loop_G3}
\end{figure}
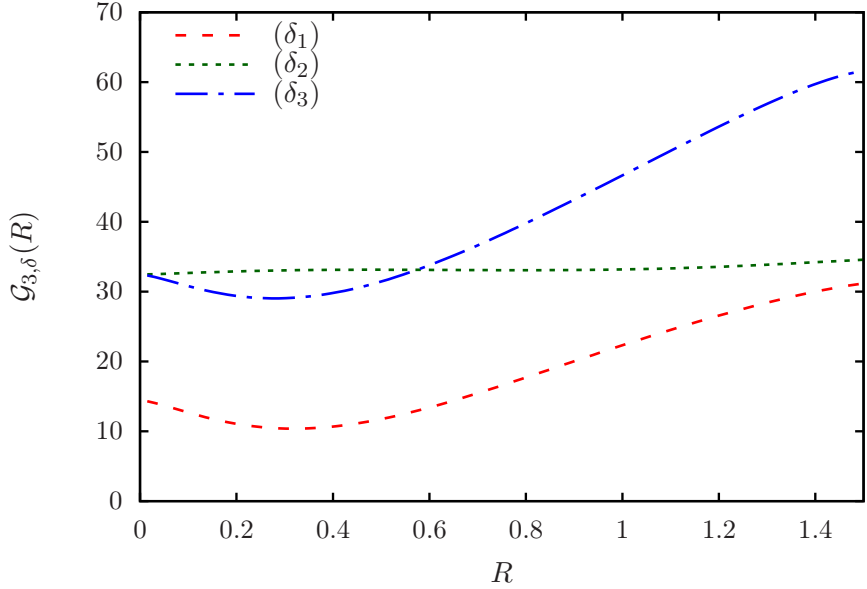

The 3-loops NGLs integral is given in an identical form to that presented in the $e^+ e^-$ case \cite{Khelifa-Kerfa:2015mma} (Eq. (3.10)). That is, 
\begin{align}\label{eq:3loop_S3_Integral}
 \cS_{3,\delta}(\rho) = - \int_{\xi_1>\xi_2>\xi_3} \d\Pi_{123}\, \Theta_1^\out \Theta_3^\inn \left(\Theta_2^\inn \oW_{123,\delta}^{\R\V\R} + \Theta_2^\out \left[ \oW_{123, \delta}^{\R\V\R} + \oW_{123, \delta}^{\R\R\R} \right]   \right),
\end{align}
where as before $\d\Pi_{123} = \prod_{i=1}^3 \d\Phi_i \Theta(\varrho_i - \rho)$, with the 1-loop phase space factor given in \eqref{eq:1loop_PhaseSpaceFactor}, and the eikonal amplitudes squared are given in Ref. \cite{Khelifa-Kerfa:2020nlc}. They read 
\begin{subequations}
\begin{align}\label{eq:EikAmp_3loop}
 \oW_{123, \delta}^{\R\R\R} &= \CAsq \sum_{(ij) \in \Delta_\delta} \cC_{ij} \left[\cA_{ij}^{12} \cAb_{ij}^{13} + \cB_{ij}^{123}\right] + Q_\delta \sum_{\pi_{\{ijk\}}} \left[\cG_{ij}^{k1}(2,3) + 2 \leftrightarrow 3 \right],
 \\
 \oW_{123, \delta}^{\R\V\R} &= - \CAsq \sum_{(ij) \in \Delta_\delta} \cC_{ij}\, \cA_{ij}^{12} \cAb_{ij}^{13}  - Q_\delta \sum_{\pi_{\{ijk\}}}  \left[\cG_{ij}^{k1}(2,3) + 2 \leftrightarrow 3 \right],
\end{align}
\end{subequations}
where $\cAb_{ij}^{k\ell} = \cA_{ij}^{k\ell}/w_{ij}^k$, the permutation $\pi_{\{ijk\}} = \{(ijk), (ikj), (jki) \}$, the 3-loops antenna function $\cB_{\alpha \beta}^{ijk}$ and the {\it quadruple} function $\cG_{ij}^{k\ell}$ are defined by 
\begin{subequations}
\begin{align}
 \cB_{\alpha \beta}^{ijk} &= w_{\alpha \beta}^i \left(\cA_{\alpha i}^{jk} + \cA_{i \beta}^{jk} - \cA_{\alpha \beta}^{jk}  \right),
 \label{eq:3loo_AntennaFun}
 \\
 \cG_{ij}^{k\ell}(n,m) &= w_{ij}^\ell T_{ij}^{k \ell}(n) U_{ij}^{k\ell}(m),
 \label{eq:3loop_QuadFun}
\end{align}
\end{subequations}
with $T_{ij}^{k\ell}(n) = w_{ij}^n + w_{k\ell}^n - w_{ik}^n - w_{j\ell}^n$ and $U_{ij}^{k\ell}(n) = w_{ij}^n + w_{k\ell}^n - w_{i\ell}^n - w_{jk}^n$. The quadrupole colour factor $Q_\delta$ reads 
\begin{align}\label{eq:3loop_QuadColorFactor}
 Q_{\delta_1} = Q_{\delta_2} = \CAsq(\CA - 2\CF) = \CA, \qquad Q_{\delta_3} = 6 \,\CA. 
\end{align}
Substituting all of the above expressions back into the formula of $\cS_{3, \delta}(\rho)$ \eqref{eq:3loop_S3_Integral} we find that, just like the 2-loops integral, up to single log accuracy the energy integrals factorise out to give $L^3/3!$. We then write the NGLs contribution at 3-loops in an analogous form to \eqref{eq:NGLs_2loop_FinalForm}: 
\begin{align}\label{eq:3loop_S3_FactorisedForm}
 \cS_{3, \delta}(\rho) = + \frac{1}{3!} \, \asb^3\, L^3\, \cG_{3, \delta}(R),
\end{align}
where 
\begin{align}\label{eq:3loop_G3}
 \cG_{3, \delta}(R) = \CAsq \sum_{(ij) \in \Delta_\delta} \cC_{ij} \left( \cJ_{ij}^{(1)}(R) -  \cJ_{ij}^{(2)}(R) \right) + 2 Q_\delta \sum_{(ijk) \in \pi_\delta} \cJ_{ijk}^{(3)}(R),
\end{align}
with $\pi_{\delta} = \{(ajb), (bja), (abj) \}$, the factor of $2$ in the second part of the rhs of the above equation comes from the ($2 \leftrightarrow 3$) symmetry of the integrand and the integral expressions of the various terms in \eqref{eq:3loop_G3} are given by 
\begin{subequations}
	\label{eq:3loop_J3_Integrals}
\begin{align}
 \cJ_{ij}^{(1)}(R) &= R^6 \int_0^{2\pi} \prod_{i=1}^3 \frac{\d\theta_i}{2\pi} \int_1^{\frac{\pi}{R |\sin\theta_1|}} r_1 \d r_1 \int_0^1 r_2 \d r_2 \int_0^1 r_3 \d r_3 \, \cA_{ij}^{12} \, \cAb_{ij}^{13}, 
 \\
 \cJ_{ij}^{(2)}(R) &= - R^6 \int_0^{2\pi} \prod_{i=1}^3 \frac{\d\theta_i}{2\pi} \int_1^{\frac{\pi}{R |\sin\theta_1|}} r_1 \d r_1 \int_1^{\frac{\pi}{R |\sin\theta_2|}} r_2 \d r_2 \int_0^1 r_3 \d r_3 \, \cB_{ij}^{123},
 \\
 \cJ_{ijk}^{(3)}(R) &= R^6 \int_0^{2\pi} \prod_{i=1}^3 \frac{\d\theta_i}{2\pi} \int_1^{\frac{\pi}{R |\sin\theta_1|}} r_1 \d r_1 \int_0^1 r_2 \d r_2 \int_0^1 r_3 \d r_3 \, \cG_{ij}^{k1}(2,3).
\end{align}
\end{subequations}
Following the procedure outlined at 2-loops one can find an analytical $R$-series expansion for all of the three integrals. They read 
\begin{subequations}
\begin{align}\label{eq:3loop_J1_aj-ab}
 \cJ_{aj}^{(1)} = \cJ_{bj}^{(1)} &= \zeta_3 + \frac{1}{16} R^2 + 0.005 R^4 - 0.001 R^6 - \cO(R^8),
 \\ 
  \cJ_{ab}^{(1)} &= \zeta_2 R^2 + \frac{1}{2} R^4(\ln R-1) - \frac{1}{32} R^6 - \cO(R^8).
\end{align}
\end{subequations}
for the first part, 
\begin{subequations}
	\begin{align}\label{eq:3loop_J2_aj-ab}
	\cJ_{aj}^{(2)} = \cJ_{bj}^{(2)} &= \frac{\zeta_3}{2} + \frac{1}{16} R^2 - 0.017 R^4 + 0.012 R^6 - \cO(R^8),
	\\ 
	\cJ_{ab}^{(2)} &= R^2(\ln^2 R - \ln R +0.5) - (0.652\ln R - 0.375)R^4 +  (0.051 \ln R - 0.004) R^6 -
	\notag \\ &- \cO(R^8),
	\end{align}
for the second part, and for the third part we sum up the three contributions $	\cJ_{ajb}^{(3)} + \cJ_{bja}^{(3)}+ \cJ_{abj}^{(3)}$, since the colour factor $Q_\delta$ is independent of the choice of the triplets $(ajb), (bja)$ and $(abj)$, to give a {\it cross-channel} coefficient 
\begin{align}\label{eq:3loop_J3X}
	\cJ_{X}^{(3)}  &= \frac{1}{4} R^2 + R^4 (0.125 \ln R - 0.022) - 0.023 R^6 + \cO(R^8). 
\end{align}
\end{subequations}
Substituting the various formulae above back into the form of $\cG_{3,\delta}(R)$ \eqref{eq:3loop_G3} we can deduce the expression of the latter for each channel. To this end we have:
\begin{subequations}
	\begin{align}\label{eq:3loop_G2_FinalChannel}
	\cG_{3,\delta_1}(R) &= \CF \CAsq \left[1.202 - R^2 + (0.09 - 0.5 \ln R) R^4 + 0.094 R^6   \right]+ \notag\\
	   &+ \CAcub \big[(1.645+\ln R - \ln^2 R) R^2 +( 1.402 \ln R - 0.919) R^4 -
	    \notag \\ & \hspace{1cm}  - (0.073+0.051 \ln R) R^6  \big] + \cO(R^8),
	\\
	\cG_{3,\delta_2}(R) &= \CF \CAsq \big[ (1.29 + 2\ln R -2\ln^2 R)R^2 + (1.805 \ln R -1.662) R^4  +
	\notag\\
	& \hspace{1.5cm} + (0.039 - 0.103 \ln R) R^6  \big] + \notag \\
	&+ \CAcub \big[ 1.202 + (\ln^2 R - \ln R - 0.645)R^2 + (0.838 - 0.902 \ln R) R^4 +  \notag\\ 
	& + (0.051 \ln R - 0.018 ) R^6\big] + \cO(R^8),
	\\
	\cG_{3,\delta_3}(R) &= \CAcub \big[ 1.202 + (1.145 +\ln R - \ln^2 R)R^2 + (1.152 \ln R - 0.868) R^4 -
	\notag \\
	& - (0.025 + 0.051 \ln R) R^6  \big]  + \CA \left[3 R^2 + (1.5 \ln R - 0.264) R^4 - 0.279 R^6 \right]+ \notag \\ &+ \cO(R^8).
	\end{align}
\end{subequations}
In Fig. \ref{fig:3loop_G3} we plot the above NGLs coefficients for all of the three channels. As expected, the large contribution comes from channel $(\delta_3)$ as it has a larger colour factor. Channel $(\delta_2)$, just as observed at 2-loops, has its NGLs coefficient almost constant over the entire range of values of the jet radius $R$. 
In the limit $R \to 0$ one finds for the above NGLs coefficients
\begin{align}
\lim_{R\to 0} \cG_{3,\delta_1}(R) = \CF\CAsq\, \zeta_3, \qquad 
\lim_{R\to 0} \cG_{3,\delta_2}(R) = \CAcub\, \zeta_3, \qquad 
\lim_{R\to 0} \cG_{3,\delta_3}(R) = \CAcub\, \zeta_3,
\end{align}
where $\zeta_3 = 1.202$. A similar result, to channel $(\delta_1)$, has been found in $e^+ e^-$ hemisphere mass distribution \cite{Khelifa-Kerfa:2015mma}. This confirms, at 3-loops, the boundary nature of NGLs seen at 2-loops.

\subsection{NGLs at 4-loops}

The expression of NGLs at 4-loops is, once again, analogous to that presented in \cite{Khelifa-Kerfa:2015mma} (Eqs. (3.21) and (3.22)). We write it in the form where all terms are separately finite as:
\begin{align}\label{eq:4loo_S4_Integral}
 \cS_{4, \delta} (\rho) &= - \int_{\xi_1 > \cdots > \xi_4} \d \Pi_{1234} \Theta_1^\out \Theta_4^\inn  \times  \Bigg[ 
\Theta_2^\inn \Theta_3^\inn   \oW^{\R\V\V\R}_{1234, \delta} + \notag \\
 &\;\; + \Theta_2^\inn \Theta_3^\out \left( \oW^{\R\V\V\R}_{1234, \delta} + \oW^{\R\V\R\R}_{1234, \delta} \right) 
+ \Theta_2^\out \Theta_3^\inn \left( \oW^{\R\V\V\R}_{1234, \delta} + \oW^{\R\R\V\R}_{1234, \delta} \right)
\notag \\
& + \Theta_2^\out \Theta_3^\out \left( \oW^{\R\V\V\R}_{1234, \delta} + \oW^{\R\V\R\R}_{1234, \delta} + \oW^{\R\R\V\R}_{1234, \delta} + \oW^{\R\R\R\R}_{1234, \delta}\right)
\Bigg].
\end{align}
The expressions of the various eikonal amplitudes squared in the above formula are given in \cite{Khelifa-Kerfa:2020nlc} and will not, for brevity, be explicitly repeated here. It is worth mentioning that all of the eikonal amplitudes above contain quadrupole (proportional to the quadrupole colour factor $Q_\delta$ \eqref{eq:3loop_QuadColorFactor}) terms that have some special features not seen in other terms (see Ref. \cite{Delenda:2015tbo, Khelifa-Kerfa:2020nlc} for details). The latter terms were  referred to as {\it ghost terms} in \cite{Delenda:2015tbo} and they correspond to: $\cNb_{1234}^{\R\R\R\R}, \cNb_{1234}^{\R\V\R\R}, \cNb_{1234}^{\R\R\V\R}$ and $\cNb_{1234}^{\R\V\V\R}$. Their corresponding expressions have not been written in an analytical closed form due to them being very cumbersome. Nonetheless, they may be integrated out easily. 

\begin{figure}[h!]
	\centering
	\input{Figs/G4l.tex}
	\caption{NGLs coefficients at 4-loops for all of the three Born channels.} \label{fig:4loop_G4}
\end{figure}
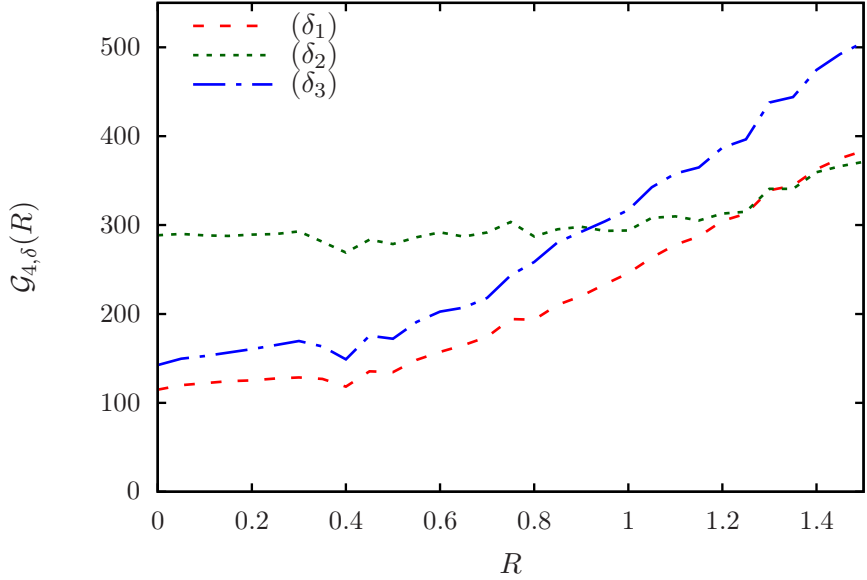

Following the 2- and 3-loops calculations we write the 4-loops NGLs contribution \eqref{eq:4loo_S4_Integral} in the form 
\begin{align}\label{eq:4loop_S4_FactorisedForm}
 \cS_{4, \delta}(\rho) = - \frac{1}{4!}\, \asb^4\, L^4\, \cG_{4, \delta}(R),  
\end{align}
where the NGLs coefficient at this order reads 
\begin{align}\label{eq:4loop_G4}
 \cG_{4, \delta}(R) &= \CAcub \sum_{(ij) \in \Delta_\delta} \cC_{ij} \left[ - \cK^{(1)}_{ij} + \cK^{(3)}_{ij} + \cK^{(5)}_{ij} + \cK^{(6)}_{ij} + \frac{Q_\delta}{\CAcub} \cK^{(7)}_{ij}   - \cK^{(9)}_{ij}   \right] - \notag \\
 &- \CA Q_\delta \left[\cK^{(2)}_{X} + \cK^{(4)}_{X} + \cK^{(8)}_{X}  \right].
\end{align}
The integral expressions of the various terms in the above equation are reported in appendix \ref{app:4loop_Integrals} (Eqs. \eqref{eq:4loop_K_Integrals}). The results of integration of each term are also presented in the same appendix. For the reasons stated in the appendix we report here the numerical values of the NGLs coefficients $\cG_{4, \delta}(R)$ for the three channels. They are plotted in Fig. \ref{fig:4loop_G4}.  
In the limit $R = 0$ we have the explicit expressions of the latter coefficients given by:  
\begin{subequations}\label{eq:4loop_G4_R=0}
\begin{align}
 \lim_{R \to 0} \cG_{4, \delta_1}(R) &= 1.08\, \CFsq \CAsq + 8.79\, \CF \CAcub  - 2.7\, \CAfour, 
 \\
 \lim_{R \to 0} \cG_{4, \delta_2}(R) &= 6.49\, \CF \CAcub  + 0.678\, \CAfour, 
 \\
 \lim_{R \to 0} \cG_{4, \delta_3}(R) &= -19.5\, \CAsq  + 3.92\, \CAfour,
\end{align}
\end{subequations}
Note that $\zeta_4 = 1.08$ and recall the $e^+ e^-$ result for the 4-loops coefficient \cite{Khelifa-Kerfa:2015mma}: $(25/8 \CF\CAcub + \CFsq \CAsq) \zeta_4 = 1.08\, \CFsq \CAsq + 3.38\, \CF \CAcub$. Thus in the limit $R \to 0$ the 4-loops NGLs coefficients in V/H+jet processes do not reduce to those of $e^+ e^-$. The source of the mismatch is the contribution $\cK_{X}^{(8)}$, and in particular the quadrupole ghost term $\cNb_{1234}^{\R\R\V\R}$. It was shown in \cite{Delenda:2015tbo} that such terms, whereby both gluons $1$ and $2$ are {\it real}, posses very peculiar features, such as breaking both mirror and Bose symmetries 	(full details are to be found in the said reference). As a confirmation to this observation, the other quadrupole contributions in Eq.  \eqref{eq:4loop_G4}, namely $\cK_{X}^{(2)}$ and $\cK_{X}^{(4)}$, tend to zero in the limit $R \to 0$, and hence do not contribute to the mismatch mentioned above. Evidently, none of them contains a term with both gluons $1$ and $2$ real.   

To assess the accuracy of our calculations up to 4-loops, we shall carry out, in the next section, comparisons to the numerical all-orders large-N$_c$ results presented in Ref. \cite{Dasgupta:2012hg}.

\section{Comparisons to all-orders results}
\label{sec:Comparisons}

\begin{figure}[t]
	\centering
	\includegraphics[scale=.5]{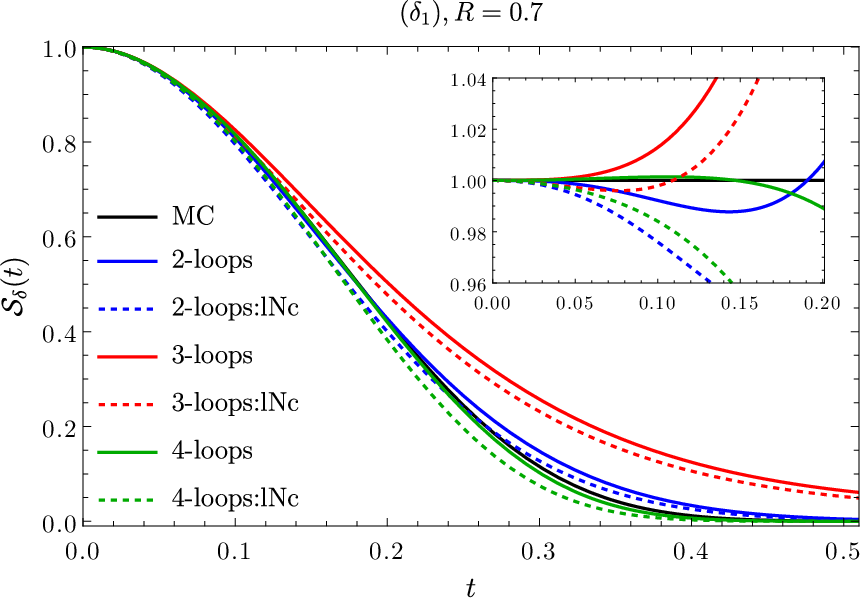}
	\includegraphics[scale=.5]{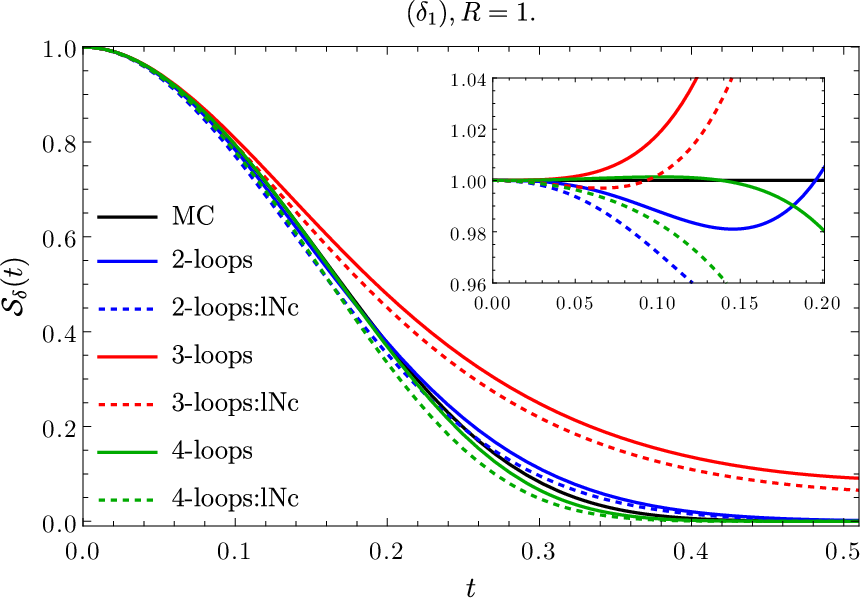}
	\vskip 1em
	\includegraphics[scale=.5]{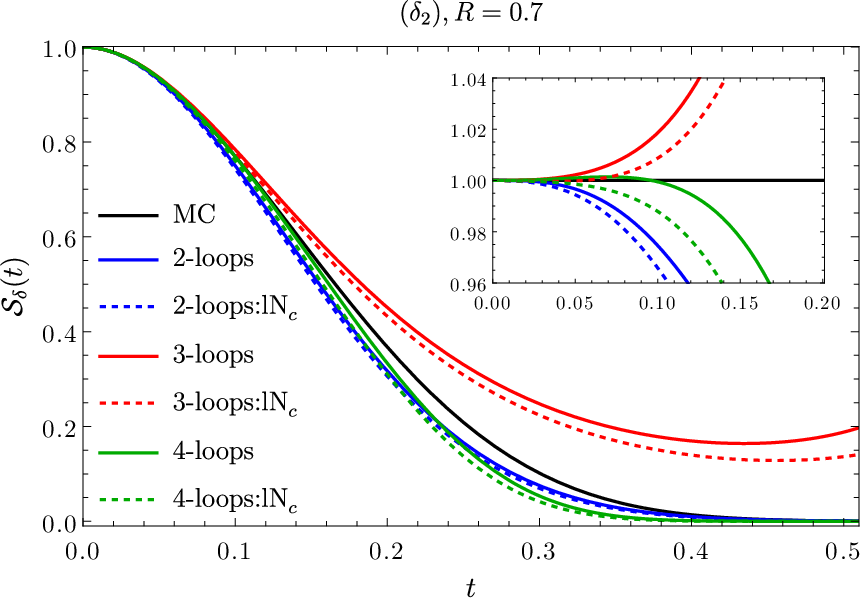}
	\includegraphics[scale=.5]{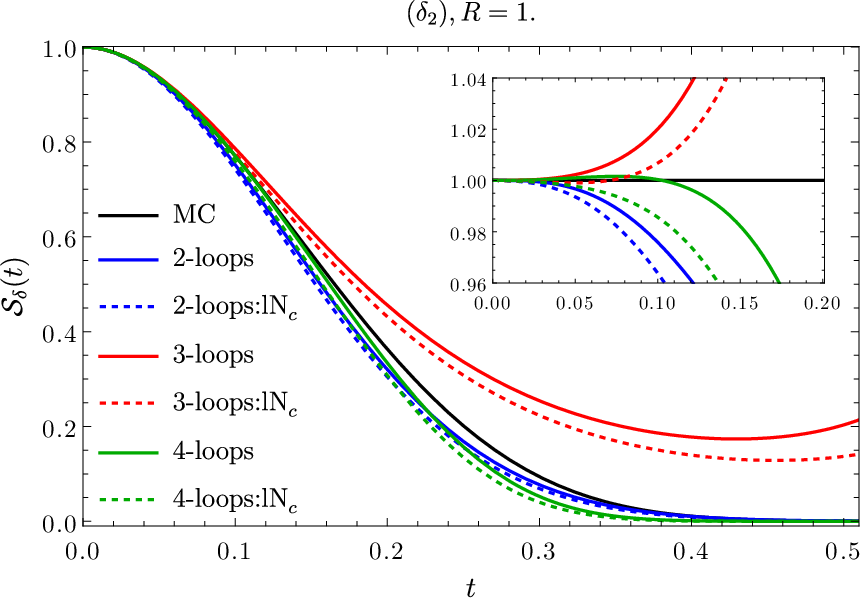}
	\vskip 1em
	\includegraphics[scale=.5]{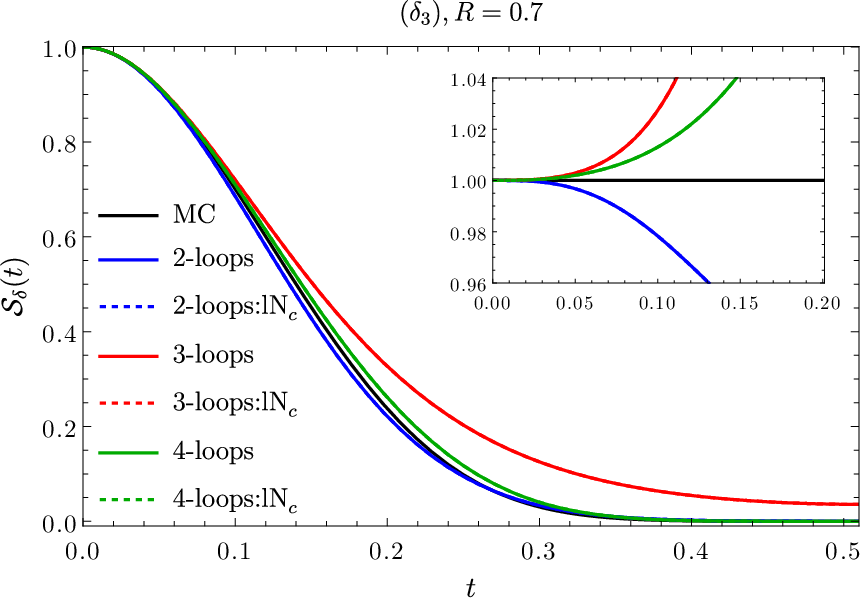}
	\includegraphics[scale=.5]{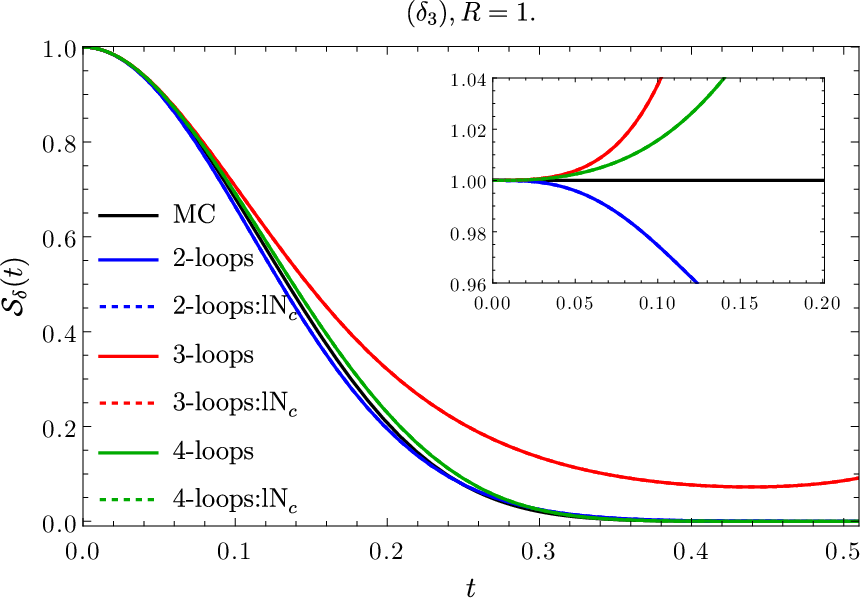}
	\caption{Comparisons between the all-orders large-N$_c$ MC result and the exponential of our analytical calculations \eqref{eq:All-loop_S_Analytic} for $R = 0.7$ (left) and $R = 1.0$ (right), for all three Born channels. Solid lines represent full colour contributions while dashed lines (with the abbreviation ``lNc" for large-N$_c$) represent large-N$_c$ contributions only.} \label{fig:St_MC_exp}
\end{figure}

\begin{figure}[t]
	\centering
	\includegraphics[scale=.5]{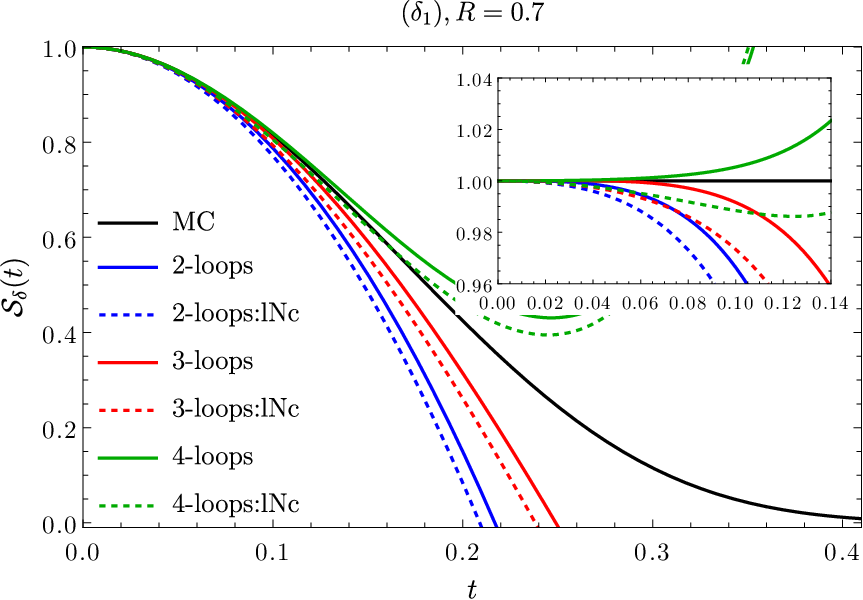}
	\includegraphics[scale=.5]{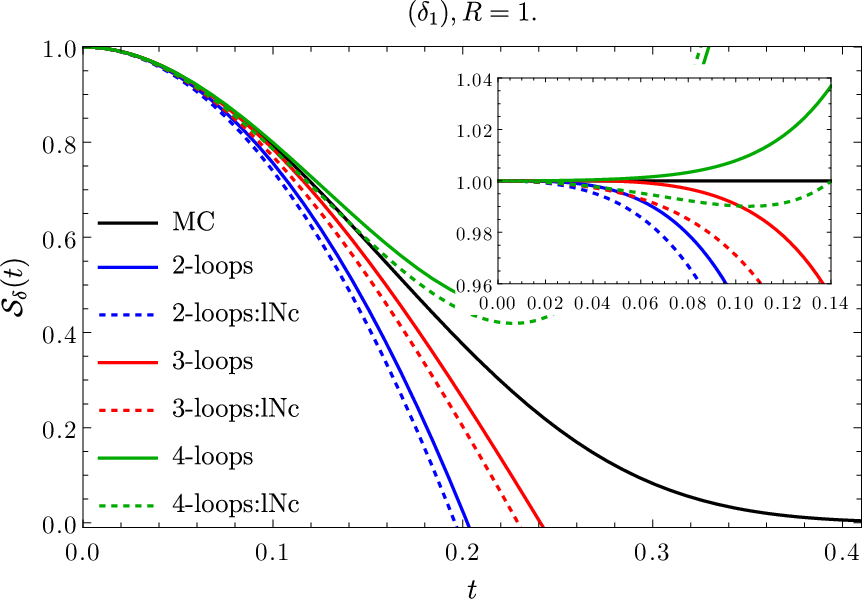}
	\vskip 1em 
	\includegraphics[scale=.5]{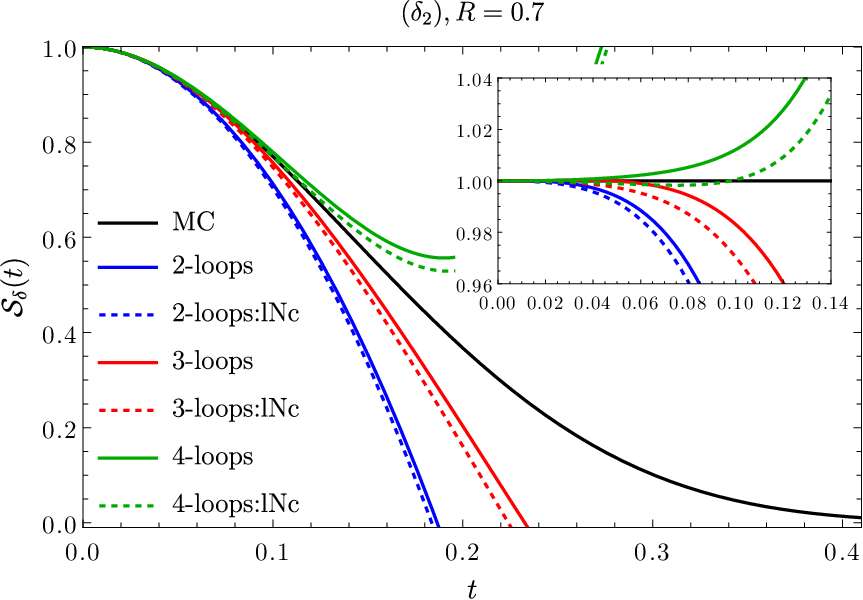}
	\includegraphics[scale=.5]{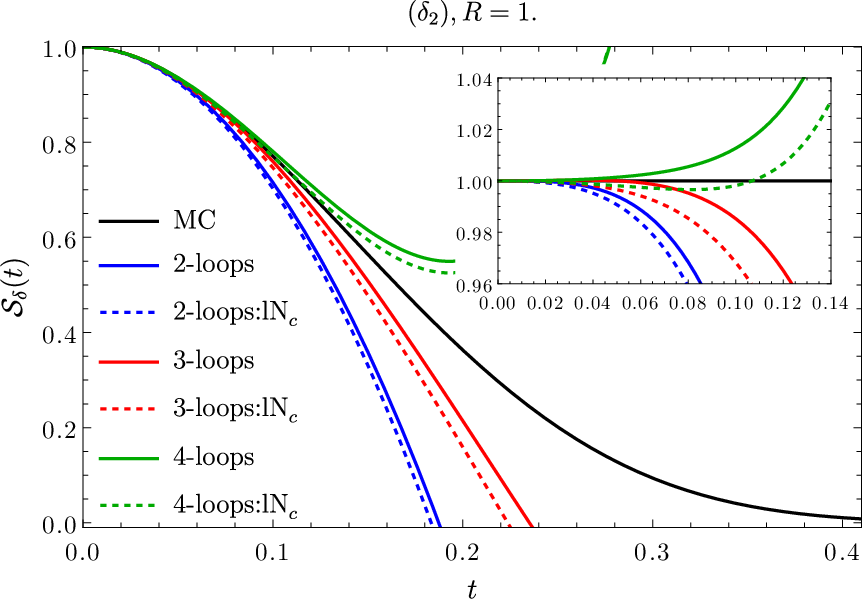}
	\vskip 1em
	\includegraphics[scale=.5]{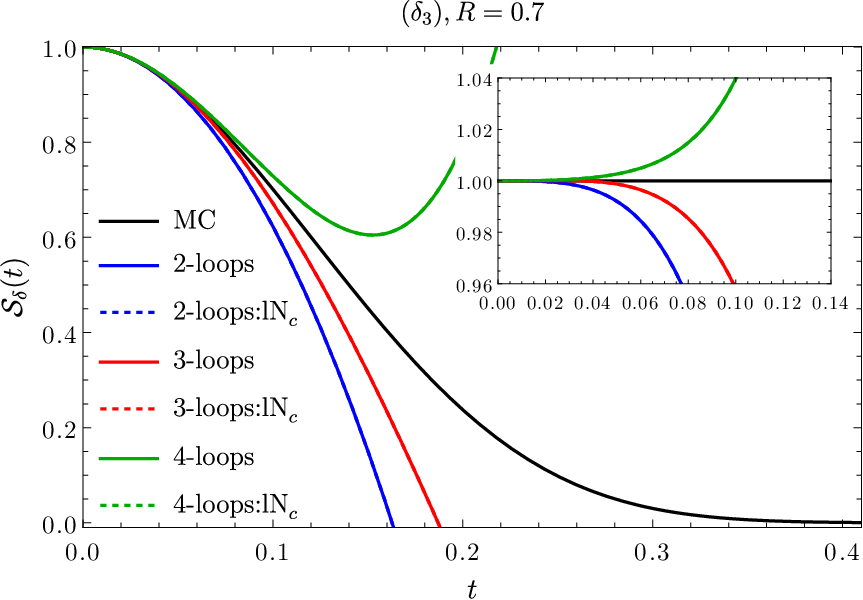}
	\includegraphics[scale=.5]{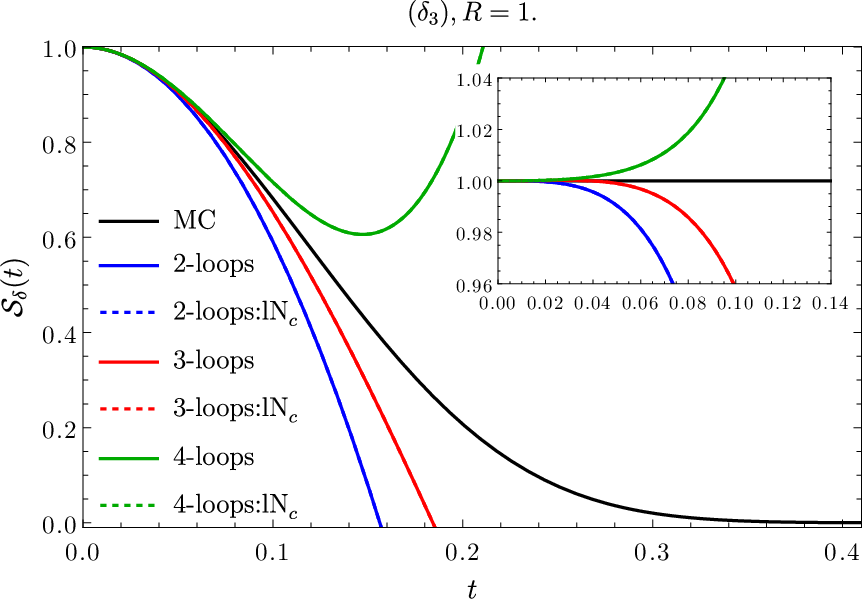}
	\caption{Comparisons between the all-orders large-N$_c$ MC result and the power series expansion of the exponential form \eqref{eq:All-loop_S_Analytic} for $R = 0.7$ (left) and $R = 1.0$ (right), for all three Born channels. Solid lines represent full colour contributions while dashed lines (with the abbreviation ``lNc" for large-N$_c$) represent large-N$_c$ contributions only.}  \label{fig:St_MC_ser}
\end{figure}

NGLs at hadron colliders have only been resummed numerically at large-N$_c$ and in the anti-k$_t$ jet algorithm using the MC code of \cite{Dasgupta:2001sh}, as reported in our previous work \cite{Dasgupta:2012hg}. In the latter reference we used the following parametrisation, which includes the full contribution at 2-loops:
\begin{align}\label{eq:All-loop_St_MC}
 \cS_{\delta}^{\MC}(t) = \exp\left[ -\CA \sum_{(ij) \in \Delta_\delta} \cC_{ij}\, \cI_{ij}\, f_{ij}(t)   \right],
\end{align}  
where $\cI_{ij}$ are the 2-loops NGLs coefficients reported in Eqs. \eqref{eq:NGLs_2loop_Iaj} and \eqref{eq:NGLs_2loop_Iab} and 
\begin{align}\label{eq:All-loop_fij-t}
f_{ij}(t) = \frac{1 + (\lambda_{ij} t)^2  }{1 + (\sigma_{ij} t)^{\gamma_{ij}}}\, t^2 , \qquad 
t = -\frac{1}{4 \pi \beta_0} \ln\left(1 - 2 \as(p_t) \beta_0\, L\right).
\end{align}
Before proceeding further we note that according to the definitions of Ref. \cite{Dasgupta:2012hg} the colour factors $\cC_{ij}$ there equal half their values here, and $\cI_{ij}$ there equal 4 times their values here. 
The functional form \eqref{eq:All-loop_St_MC} is then fitted to the output of the MC code for each of the dipoles $(ij)$ and the values of the fitting parameters $\lambda_{ij}, \sigma_{ij}$ and $\gamma_{ij}$ are given in appendix C of \cite{Dasgupta:2012hg}. We shall only consider two values of $R$, namely $0.7$ and $1.0$, for our comparisons. For the said values the fitting parameters are: 
\begin{subequations}\label{eq:All-loop_FittingVals}
\begin{align}
R = 0.7: & \qquad \lambda_{aj}=\lambda_{bj} = 0.79 \CA, \sigma_{aj}= \sigma_{bj} = 0.82 \CA, \gamma_{aj}=\gamma_{bj} = 1.33, \notag\\
             & \qquad \lambda_{ab}= 0.96 \CA, \sigma_{ab} = 0.29 \CA, \gamma_{ab} = 1.33, 
             \\
R = 1.0: & \qquad \lambda_{aj}=\lambda_{bj} = 0.86 \CA, \sigma_{aj}= \sigma_{bj} = 0.85 \CA, \gamma_{aj}=\gamma_{bj} = 1.33, \notag\\
& \qquad \lambda_{ab}= 1.24 \CA, \sigma_{ab} = 0.80 \CA, \gamma_{ab} = 1.33.        
\end{align}
\end{subequations}
Note that at fixed order the evolution parameter $t$ in Eq. \eqref{eq:All-loop_fij-t} reduces to $\asb L/2$. We compare \eqref{eq:All-loop_St_MC} to the exponential of our analytical NGLs results. That is: 
\begin{align}\label{eq:All-loop_S_Analytic}
 \cS_\delta (t) = \exp\left[ -\frac{1}{2!} \, \cG_{2, \delta}(R) (2t)^2 + \frac{1}{3!} \, \cG_{3, \delta}(R) (2t)^3  -\frac{1}{4!} \, \cG_{4, \delta}(R) (2t)^4  \right], 
\end{align}  
In Fig. \ref{fig:St_MC_exp} we plot the NGLs resummed factor \eqref{eq:All-loop_St_MC} along with our analytical exponential factor \eqref{eq:All-loop_S_Analytic} for $R = 0.7$ and $R = 1.0$ in all three Born channels. 
The solid lines represent the full expression \eqref{eq:All-loop_S_Analytic} including finite-N$_c$ contributions, whilst dashed lines represent the expression \eqref{eq:All-loop_S_Analytic} in the large-N$_c$ limit. i.e., the limit $\CF \to \CA/2$. 
The label ``2-loops'' means that we truncate the expression in the exponent \eqref{eq:All-loop_S_Analytic} at $t^2$, ``3-loops'' at $t^3$ and so on. Moreover, Fig. \ref{fig:St_MC_ser} is analogous to Fig. \ref{fig:St_MC_exp} except that it corresponds to comparisons of \eqref{eq:All-loop_St_MC} and the power series expansion of \eqref{eq:All-loop_S_Analytic}.

Concerning the exponential form (Fig. \ref{fig:St_MC_exp}), overall, the 2- and 4-loops curves seem to better represent the all-orders MC curve for the whole range of $t$. Zooming in one observes that, for smaller values of $t$, the 4-loops curve has a complete overlapping with the MC result over a slightly larger range of $t$ than the 2- and 3-loops curves, particularly for channels $(\delta_1)$ and $(\delta_2)$. A  difference of less than $10\%$ is achieved for values of $t$ reaching out to about $0.2$, or even higher for some cases, for both jet radii and for all Born channels, with the 4-loops performing better in most cases. The dependence on the jet radius $R$ seems not to be significant. Moreover, the 2-loops exponential factor seems to approximate the all-orders to a good extent for all channels and all jet radii (this observation has been reported previously \cite{Banfi:2010pa, Khelifa-Kerfa:2015mma}). 

\begin{figure}[t]
	\centering
	\includegraphics[scale=.51]{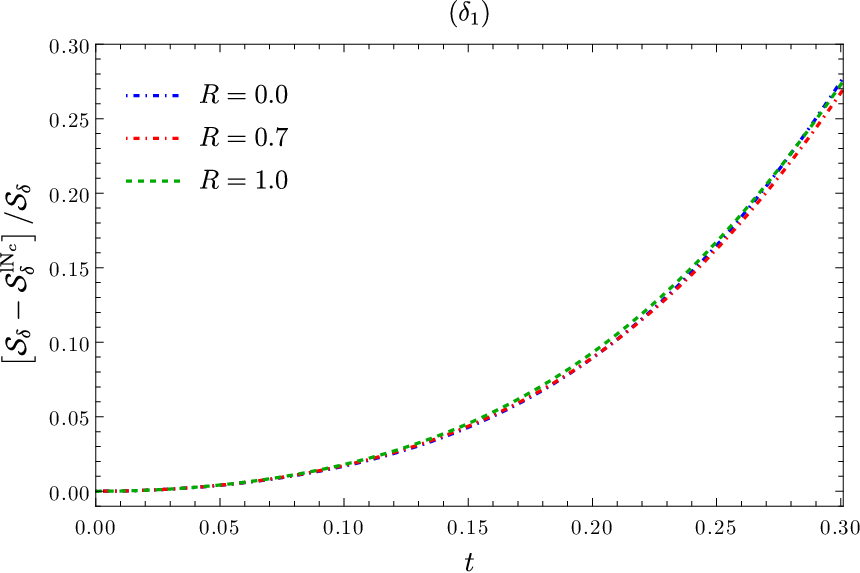}
	\includegraphics[scale=.51]{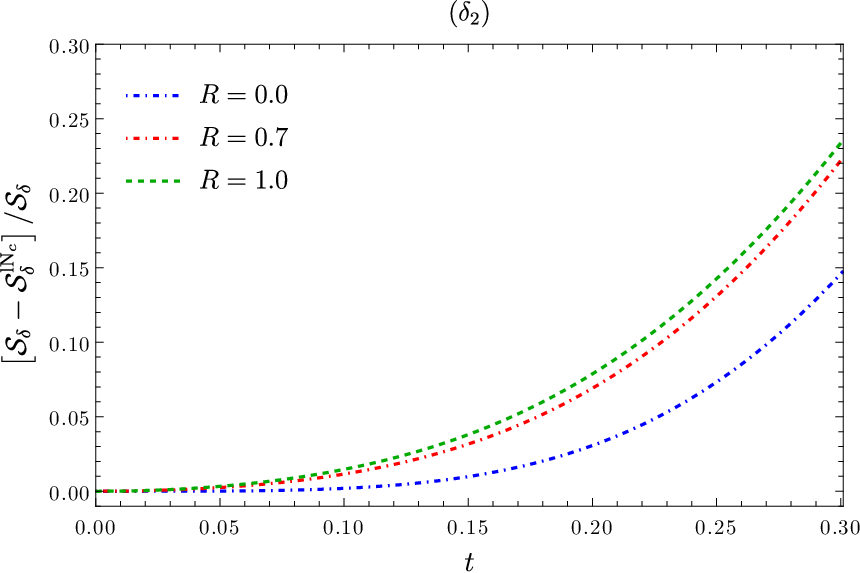}
	\caption{Comparisons between the analytical form factor \eqref{eq:All-loop_S_Analytic} with $(\cS_{\delta})$ and without $(\cS_{\delta}^{\text{lN}_c})$ the finite-N$_c$ contributions for all three jet radii, and for (left) channel $(\delta_1)$, and (right) channel $(\delta_2)$. }  \label{fig:LargeNc_St}
\end{figure}

Furthermore, curves which include the finite-N$_c$ contributions perform better than those that do not include them, especially for the 2- and 4-loops. This is true for channels $(\delta_1)$ and $(\delta_2)$ where quarks are present. The pure gluonic channel $(\delta_3)$ exhibits no difference between the two cases. In Fig. \ref{fig:LargeNc_St} we plot the percentage difference for the exponential form factor \eqref{eq:All-loop_S_Analytic} with and without finite-N$_c$ contribution for all three jet radii and for the aforementioned two channels. The difference ranges from about $0.4\%$ at $t \sim 0.05$ up to more than $24\%$ at $t \sim 0.3$ for both channels and most jet radii.    

Concerning the power series expansion, we observe, from Fig. \ref{fig:St_MC_ser},  that the higher the truncation order the better the agreement with the all-orders MC curve. This is evident in the zoomed-in plots, again especially for the first and second channels ($(\delta_1)$ and $(\delta_2)$). Compared to the exponential form, the agreement is slightly worse and over a smaller range of $t$. We shall see in the next section that this behaviour might be significantly improved by applying some carefully chosen conformal transformations. As for the comparison between the full and large-N$_c$ curves, similar features to those highlighted in the exponential case are observed.  

It has been shown by $e^+ e^-$ fixed-order calculations of NGLs up to 12-loops in the large-N$_c$ limit \cite{12loop} as well as arguments based on the analytic structure of the large-N$_c$ Banfi-Marchesini-Smye (BMS) equation \cite{Banfi:2002hw} that NGLs series has a radius of convergence of about $1$ (in the log defined in Ref. \cite{Schwartz:2014wha}) \cite{Larkoski:2016zzc}. This corresponds, in our calculations, to $t \leq 1/(2N_c) \approx 0.17$. Thus, NGLs series distribution in V/H+jet processes at hadron colliders seems to have the same gross features seen in $e^+ e^-$ collisions.

\subsection{Conformal transformation}

In order to improve the convergence of the analytical power series expansion  (towards the all-orders solution) and extend its domain of analyticity we follow Ref. \cite{Larkoski:2016zzc}, whose method is based on the original work of \cite{Altarelli:1994vz, Caprini:2000js, Caprini:2010ir, Zinn-Justin:2002ecy}, and introduce the following conformal transformations:
\begin{align}\label{eq:ConformalTrans}
   t \to u(t) = \left\{ 
   \begin{array}{cc}
     \frac{\sqrt{1+t}-1}{\sqrt{1+t}+1}, \\
     \log(1+t).
   \end{array}  \right. 
\end{align}
\begin{figure}[t]
	\centering
	\includegraphics[scale=.5]{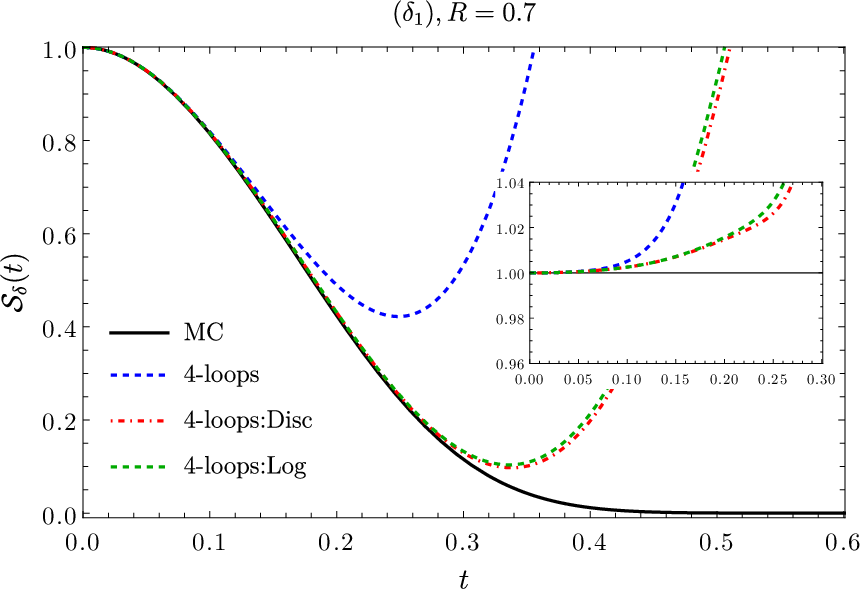}
	\includegraphics[scale=.5]{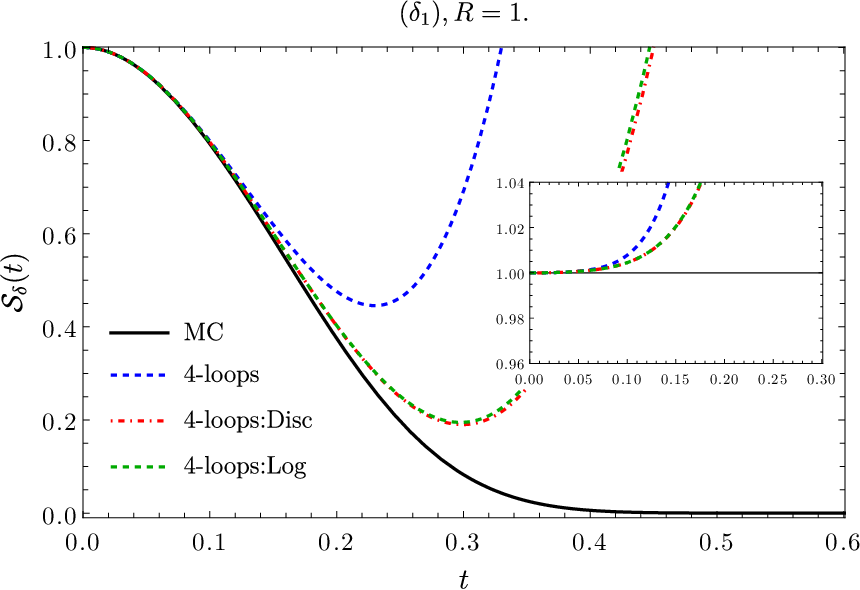}
	\vskip 1em 
	\includegraphics[scale=.5]{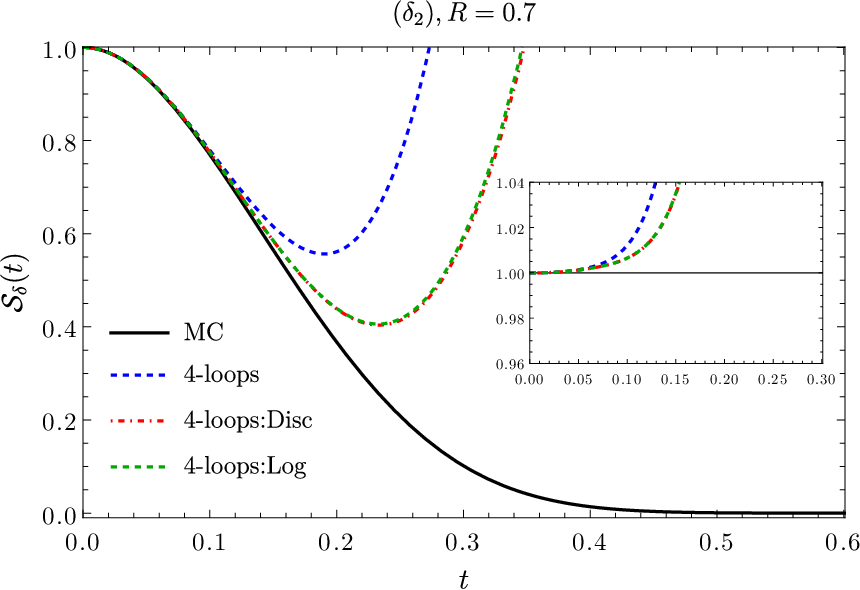}
	\includegraphics[scale=.5]{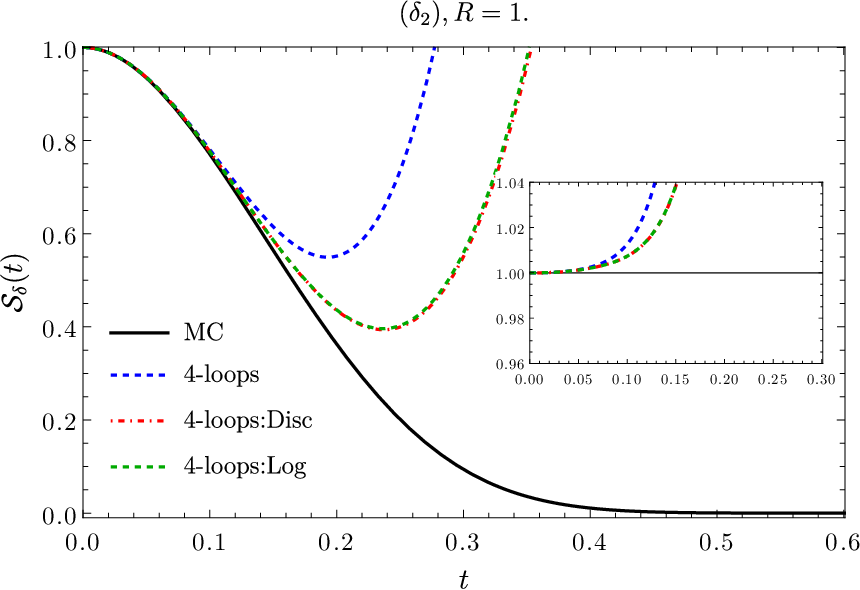}
	\vskip 1em 
	\includegraphics[scale=.5]{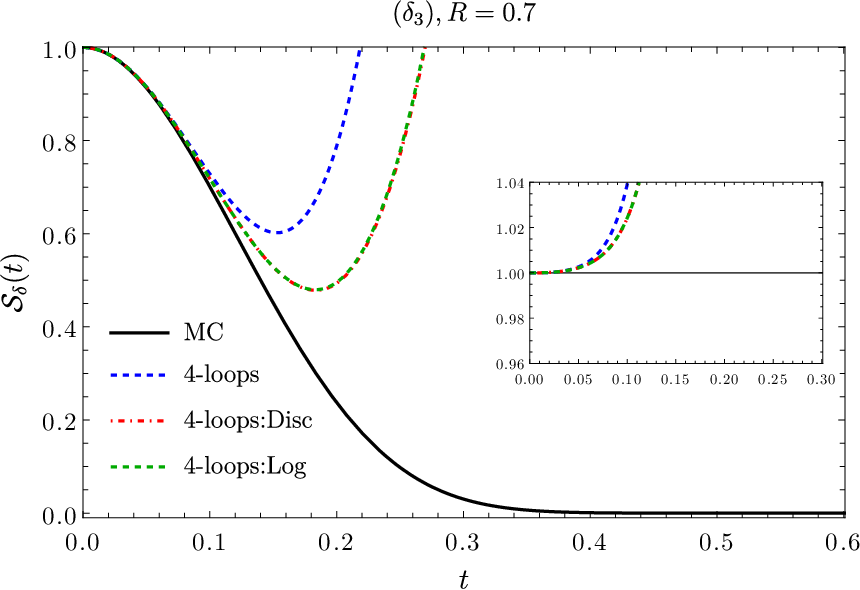}
	\includegraphics[scale=.5]{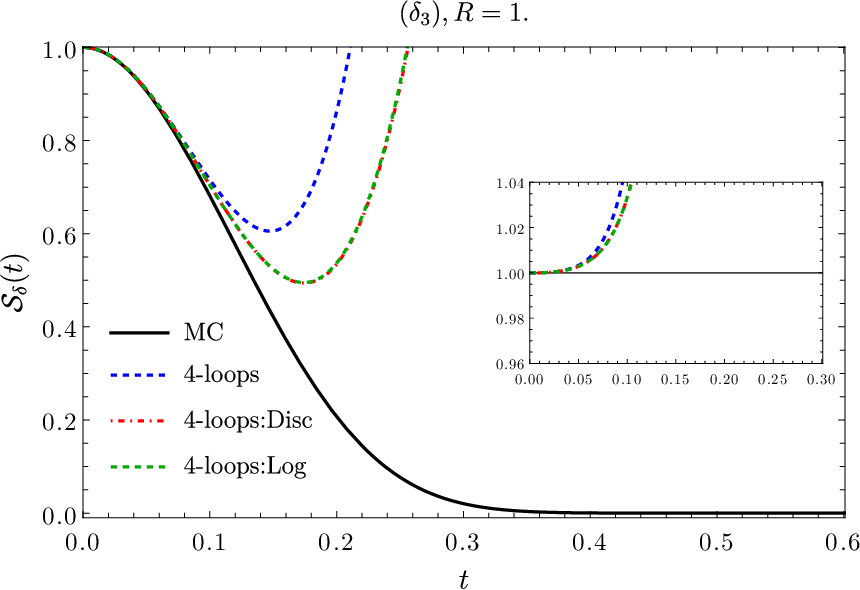}
	\vskip 1em 
	\caption{Comparisons between the all-orders large-N$_c$ MC result and the series expansion up to 4-loops with and without conformal improvement for both $R=0.7$ and $R=1.0$ and for all Born channels.} \label{fig:St_MC_exp_ConfMap}
\end{figure}
The transformation in the first line, referred to as {\it disc mapping}, maps the $t$ plane to a disc, while the other transformation, referred to as {\it log mapping}, was derived in \cite{Larkoski:2016zzc} based on the dressed gluon expansion technique. Notice that $u(0) = 0$ for both transformations. This is a necessary condition for conformal mappings. Moreover, from Eq. \eqref{eq:ConformalTrans} we infer the inverse relations: $ t= 4u/(1-u)^2 $ and $ t = e^u - 1$. To apply the said transformations we first substitute for the parameter $t$ in the expansion of Eq. \eqref{eq:All-loop_S_Analytic} by the latter expressions, power expand in terms of $u$ and then substitute back for $u$ the expressions in \eqref{eq:ConformalTrans}. 

In Fig. \ref{fig:St_MC_exp_ConfMap} we compare the NGLs series expansion up to  4-loops with and without conformal improvements against the numerical MC distribution. We observe that both disc and log mappings perform equally well in improving the convergence of the power series. For instance, for channel $(\delta_1)$ excellent agreement is achieved for up to $t \sim 0.3$ and $t \sim 0.2$ for $R = 0.7$ and $R = 1.0$, respectively. This is almost the double of that seen for the expansion without conformal improvement. Nevertheless, the conformal improvement seems to vary noticeably with the Born channel and slightly with the jet radius, performing worst for channel $(\delta_3)$ and $R=1.0$. It is worth noting, as stated in \cite{Larkoski:2016zzc}, that conformal mappings are a form of resummation only captured in a purely algebraic form, hence they allow for a larger radius of convergence when compared to fixed-order expansions. 

\begin{figure}[h!]
	\centering
	\includegraphics[scale=.5]{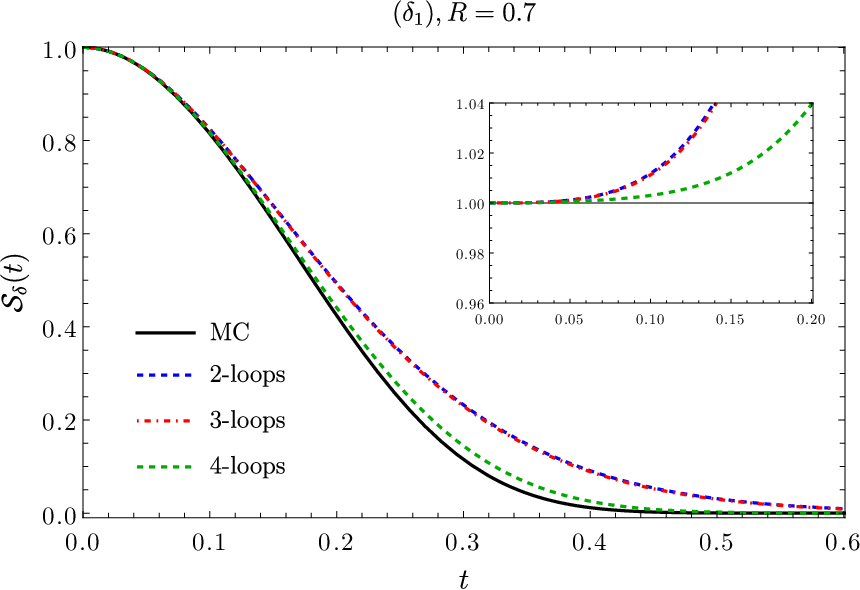}
	\includegraphics[scale=.5]{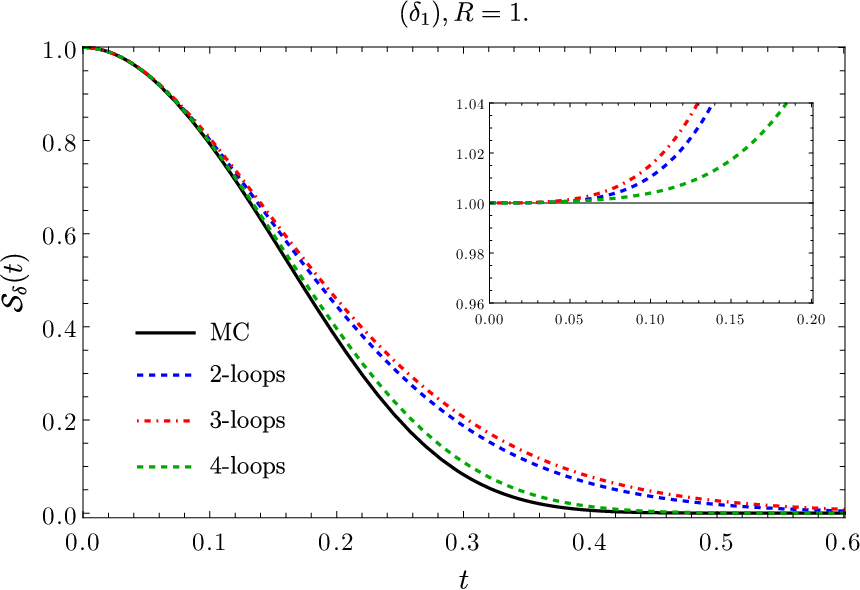}
	\vskip 1em 
	\includegraphics[scale=.5]{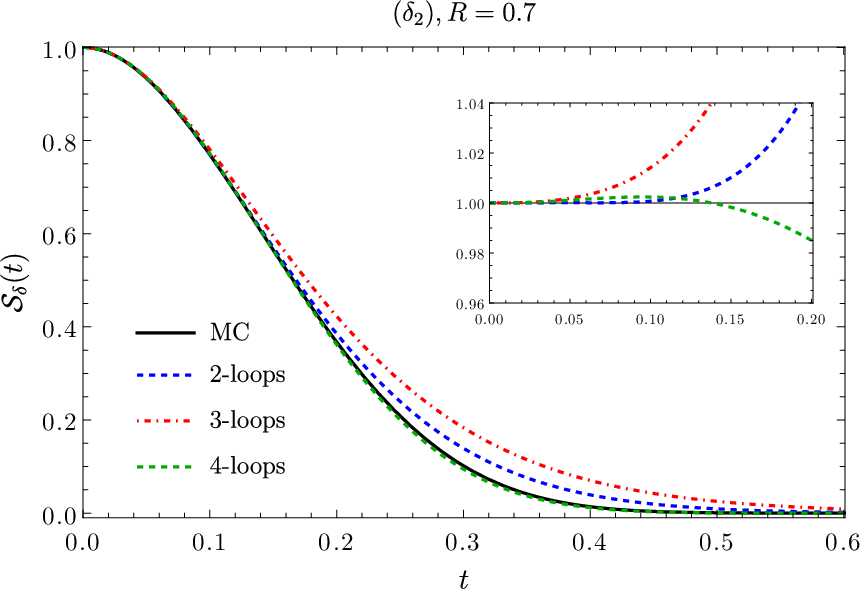}
	\includegraphics[scale=.5]{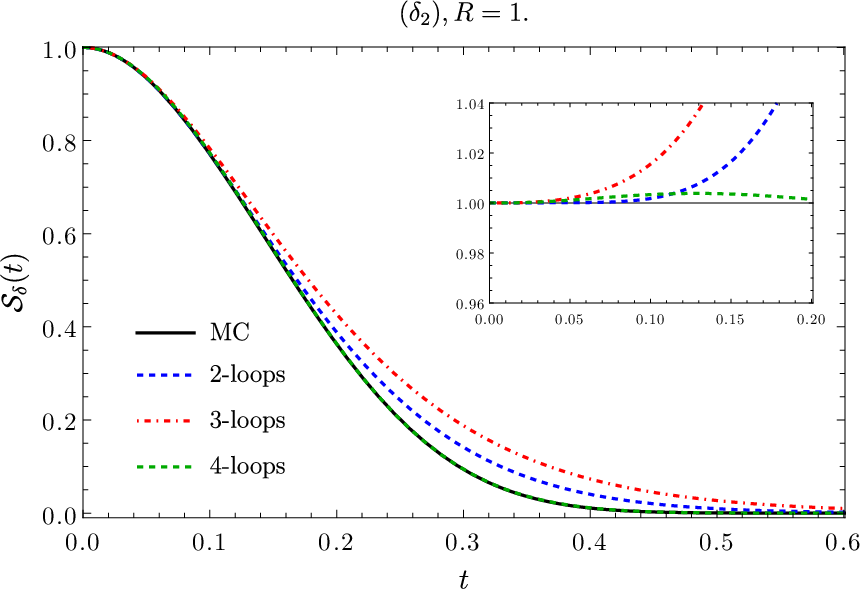}
	\vskip 1em 
	\includegraphics[scale=.5]{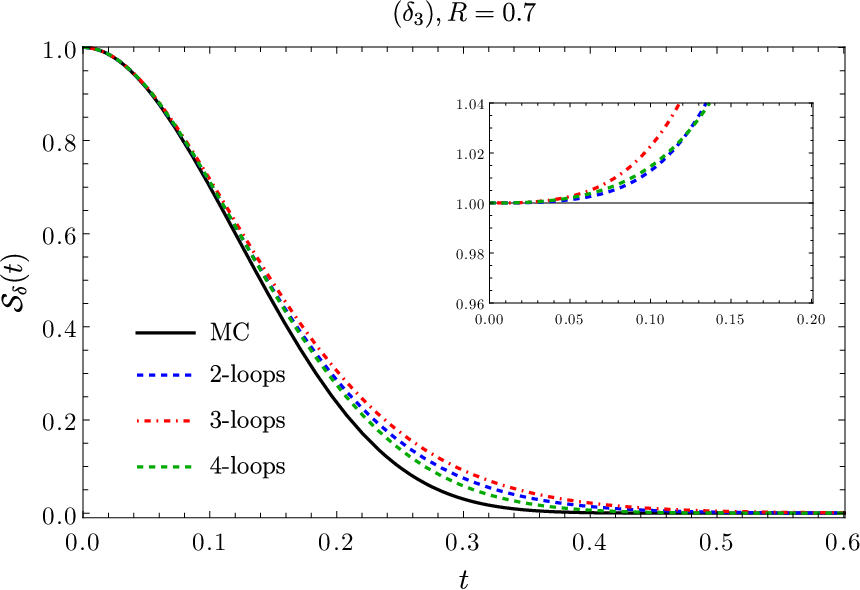}
	\includegraphics[scale=.5]{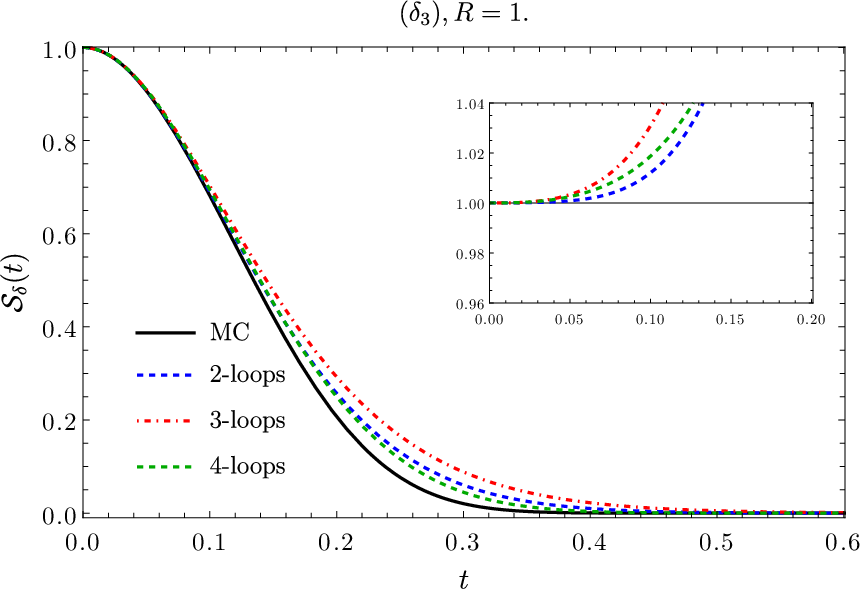}
	\vskip 1em 
	\caption{Comparisons between the all-orders large-N$_c$ MC result and the conformally improved logarithm of the analytical series expansion of \eqref{eq:All-loop_FittingVals} for both $R = 0.7$ and $R = 1.0$ and for all Born channels, } \label{fig:St_MC_LogS_ConfMap}
\end{figure}

Another possible way to further improve the convergence of the series expansion is to apply the conformal mapping to the logarithm of the expansion distribution and then exponentiate it again (as the exponentiation cancels out the logarithm) \cite{Larkoski:2016zzc}. In Fig. \ref{fig:St_MC_LogS_ConfMap} we show comparisons between the all-orders MC distribution and the said conformal improvement of the logarithm of the series expansion. Excellent agreement is evident for quite large values of $t$ extending to about $t \sim 0.4$ and beyond for some cases. This is a significant improvement compared to the original series expansion shown in Fig. \ref{fig:St_MC_ser}. It is comparable to, or even better in many cases than, the exponential form plotted in Fig. \ref{fig:St_MC_exp}.   

\section{Conclusion}
\label{sec:Conclusion}

In this paper, we have computed, for the first time in the literature, the distribution of NGLs up to 4-loops in processes involving a Higgs/vector boson + a single hard jet at hadron colliders. Our calculations focus on the invariant mass of the leading hard jet using the anti-k$_t$ jet algorithm. They are valid in the eikonal (soft) limit with strong-energy ordering, accurate to single logarithmic accuracy, and incorporate the full colour (finite-N$_c$) and jet radius $R$ dependencies of the various NGL coefficients.

Calculations at 2- and 3-loops have been performed analytically, including explicit $R$ dependence. We demonstrate that as $R \to 0$, NGL coefficients at these loop orders converge to those observed in the $e^+ e^-$ hemisphere mass distribution. However, computation of 4-loop coefficients involves partial numerical methods due to complexities arising from quadrupole ghost terms in squared eikonal amplitudes. These terms exhibit distinctive features, leading to the non-trivial observation that the $e^+ e^-$ results are not fully recovered in the $R \to 0$ limit for 4-loop coefficients, contrasting with 2- and 3-loops results.

Comparison with all-orders large-N$_c$ numerical results reveals that incorporating higher-loop NGL coefficients improves agreement with numerical MC simulations, particularly for small values of the evolution parameter $t$ (around $t \lesssim 0.15$). Notably, 2- and 4-loop results outperform the 3-loop results in capturing the overall shape of the MC curve. The performance is less satisfactory for the series expansion of the NGLs resummed form factor, indicating a convergence radius around $t \sim 0.17$, consistent with previous findings. One way to improve this convergence is to use conformal transformations, which are a type of resummation. The resultant conformally improved analytical distributions show excellent agreement with the all-orders solution over a large range of values of the evolution parameter $t$, thus proving the importance of such transformations for accurate studies. 

Furthermore, we have found that the overall features observed for NGLs calculations in $e^+ e^-$ collisions extend to hadronic V/H+jet processes. This includes, for instance, the significance of finite-N$_c$ contributions and the impact of jet radius in the phenomenological studies of jet shapes.  
Whether the same also applies for hadronic $2 \to 2$ scattering processes will be investigated in the near future.

\subsection*{Acknowledgement}

I would like to thank Y. Delenda for valuable discussions, insightful comments, and for reviewing the manuscript.

\appendix
\section{4-loops calculations}
\label{app:4loop_Integrals}

In this appendix we present the details of the 4-loops calculations. The integral expressions of the terms mentioned in Eq. \eqref{eq:4loop_G4} are given by:

\begin{subequations}\label{eq:4loop_K_Integrals}
\begin{align}
 \cK_{\alpha \beta}^{(1)} &= R^8 \int_0^{2\pi} \prod_{i=1}^4 \frac{\d\theta_i}{2\pi} \, \int_1^{\frac{\pi}{R|\sin\theta_1|}} r_1 \d r_1 \int_0^1 \prod_{j=2}^{4} r_j \d r_j  \, \cA_{\alpha \beta}^{12} \cAb_{\alpha \beta}^{13} \cAb_{\alpha \beta}^{14},
 \\
 \cK_{X}^{(2)} &= R^8 \int_0^{2\pi} \prod_{i=1}^4 \frac{\d\theta_i}{2\pi} \, \int_1^{\frac{\pi}{R|\sin\theta_1|}} r_1 \d r_1 \int_0^1 \prod_{j=2}^{4} r_j \d r_j \, \cNb_{1234}^{\R\V\V\R},
 \end{align}
 \begin{align}
 \cK_{\alpha \beta}^{(3)} &= R^8 \int_0^{2\pi} \prod_{i=1}^4 \frac{\d\theta_i}{2\pi} \, \prod_{j=1,3} \int_1^{\frac{\pi}{R|\sin\theta_j|}} r_j \d r_j \int_0^1 \prod_{\ell=2,4} r_\ell \d r_\ell \, \cA_{\alpha \beta}^{12} \cBb_{\alpha \beta}^{134},
 \\
  \cK_{X}^{(4)} &= R^8 \int_0^{2\pi} \prod_{i=1}^4 \frac{\d\theta_i}{2\pi} \, \prod_{j=1,3} \int_1^{\frac{\pi}{R|\sin\theta_j|}} r_j \d r_j \int_0^1 \prod_{\ell=2,4} r_\ell \d r_\ell \, \left(\cNb_{1234}^{\R\V\V\R} +  \cNb_{1234}^{\R\V\R\R}\right), 
  \end{align}
  \begin{align}
 \cK_{\alpha \beta}^{(5)} &= R^8 \int_0^{2\pi} \prod_{i=1}^4 \frac{\d\theta_i}{2\pi} \, \prod_{j=1,2} \int_1^{\frac{\pi}{R|\sin\theta_j|}} r_j \d r_j \int_0^1 \prod_{\ell=3,4} r_\ell \d r_\ell \, \left(\cA_{\alpha \beta}^{13} \cBb_{\alpha \beta}^{124} + \cA_{\alpha \beta}^{14} \cBb_{\alpha \beta}^{123} \right),
 \\
 \cK_{\alpha \beta}^{(6)} &= R^8 \int_0^{2\pi} \prod_{i=1}^4 \frac{\d\theta_i}{2\pi} \, \prod_{j=1,2} \int_1^{\frac{\pi}{R|\sin\theta_j|}} r_j \d r_j \int_0^1 \prod_{\ell=3,4} r_\ell \d r_\ell \, \fA_{\alpha \beta}^{1234}, 
  \end{align}
\begin{align}
\cK_{\alpha \beta}^{(7)} &= R^8 \int_0^{2\pi} \prod_{i=1}^4 \frac{\d\theta_i}{2\pi} \, \prod_{j=1,2} \int_1^{\frac{\pi}{R|\sin\theta_j|}} r_j \d r_j \int_0^1 \prod_{\ell=3,4} r_\ell \d r_\ell \, \bA_{\alpha \beta}^{1234},
  \\
\cK_{X}^{(8)} &= R^8 \int_0^{2\pi} \prod_{i=1}^4 \frac{\d\theta_i}{2\pi} \, \prod_{j=1,2} \int_1^{\frac{\pi}{R|\sin\theta_j|}} r_j \d r_j \int_0^1 \prod_{\ell=3,4} r_\ell \d r_\ell \, \left(\cNb_{1234}^{\R\V\V\R} +  \cNb_{1234}^{\R\R\V\R}\right), 
\\
\cK_{\alpha \beta}^{(9)} &= R^8 \int_0^{2\pi} \prod_{i=1}^4 \frac{\d\theta_i}{2\pi} \, \prod_{j=1,2,3} \int_1^{\frac{\pi}{R|\sin\theta_j|}} r_j \d r_j \int_0^1 r_4 \d r_4 \, \cC_{\alpha \beta}^{1234},
\end{align}
\end{subequations}
where the various angular functions are given in Ref. \cite{Khelifa-Kerfa:2020nlc}. Since the integration of some the above integrals, particularly the quadrupole ghost terms and the terms involving the jet parton as one of the dipole legs, have proven to be analytically formidable we shall confine ourselves to just provide the numerical values of the integrations. They are reported in Table \ref{tab:4loop_NumericalVals}. Notice that $\cK_{(ij)}^{(5)} = 2 \cK_{(ij)}^{(3)}$ and thus will not be shown explicitly in the latter table.
 
\begin{table}
\makebox[\linewidth]{	
\begin{tabularx}{6.93in}{|c | c | c | c | c | c | c | c | c | c | c | c | c | c |}
	\cline{2-14}
\multicolumn{1}{c|}{ }	& \multicolumn{2}{|c|}{$\cK_{ij}^{(1)}$} & \multirow{2}{*}{$\cK_{X}^{(2)}$} &  \multicolumn{2}{|c|}{$\cK_{ij}^{(3)}$}     & \multirow{2}{*}{$\cK_{X}^{(4)}$}  &  \multicolumn{2}{|c|}{$\cK_{ij}^{(6)}$}     &  
	\multicolumn{2}{|c|}{$\cK_{ij}^{(7)}$}     & \multirow{2}{*}{$\cK_{X}^{(8)}$} &  \multicolumn{2}{|c|}{$\cK_{ij}^{(9)}$} \\
	\cline{1-3} \cline{5-6} \cline{8-11} \cline{13-14}
	$R$ &  $\cK_{aj}^{(1)}$ & $\cK_{ab}^{(1)}$ &  & $\cK_{aj}^{(3)}$ & $\cK_{ab}^{(3)}$ & & $\cK_{aj}^{(6)}$ & $\cK_{ab}^{(6)}$ & $\cK_{aj}^{(7)}$ & $\cK_{ab}^{(7)}$ &  &   $\cK_{aj}^{(9)}$ & $\cK_{ab}^{(9)}$ \\
	\hline
$0.$ & $3.25$ & $0.$ & $0.$ & $0.474$ & $0.$ & $0.$ & $1.15$ & $0.$ & $0.271$ & $0.$ & $2.7$ & $1.29$ & $0.$ \\
$0.1$ & $3.25$ & $0.032$ & $-0.015$ & $0.461$ & $0.017$ & $0.01$ & $1.15$ & $0.076$ & $0.275$ & $0.001$ & $2.64$ & $1.25$ & $0.177$ \\
$0.2$ & $3.25$ & $0.129$ & $-0.059$ & $0.462$ & $0.060$ & $0.035$ & $1.15$ & $0.212$ & $0.29$ & $0.010$ & $2.51$ & $1.25$ & $0.373$ \\
$0.3$ & $3.25$ & $0.287$ & $-0.131$ & $0.464$ & $0.12$ & $0.069$ & $1.15$ & $0.363$ & $0.315$ & $0.037$ & $2.37$ & $1.27$ & $0.557$ \\
$0.4$ & $3.25$ & $0.5$ & $-0.229$ & $0.465$ & $0.187$ & $0.106$ & $1.15$ & $0.509$ & $0.348$ & $0.087$ & $2.22$ & $1.12$ & $0.689$ \\
$0.5$ & $3.26$ & $0.759$ & $-0.35$ & $0.466$ & $0.255$ & $0.147$ & $1.15$ & $0.641$ & $0.391$ & $0.165$ & $2.1$ & $1.17$ & $0.903$ \\
$0.6$ & $3.27$ & $1.06$ & $-0.492$ & $0.454$ & $0.319$ & $0.177$ & $1.15$ & $0.759$ & $0.441$ & $0.27$ & $1.99$ & $1.21$ & $1.11$ \\
$0.7$ & $3.28$ & $1.38$ & $-0.651$ & $0.456$ & $0.376$ & $0.208$ & $1.16$ & $0.859$ & $0.499$ & $0.4$ & $1.9$ & $1.21$ & $1.26$ \\
$0.8$ & $3.31$ & $1.72$ & $-0.832$ & $0.477$ & $0.433$ & $0.235$ & $1.16$ & $0.954$ & $0.268$ & $0.552$ & $1.84$ & $1.2$ & $1.44$ \\
$0.9$ & $3.34$ & $2.07$ & $-1.03$ & $0.461$ & $0.471$ & $0.266$ & $1.16$ & $1.05$ & $0.277$ & $0.719$ & $1.8$ & $1.19$ & $1.56$ \\
$1.$ & $3.4$ & $2.42$ & $-1.25$ & $0.48$ & $0.504$ & $0.285$ & $1.17$ & $1.14$ & $0.295$ & $0.899$ & $1.77$ & $1.19$ & $1.74$ \\
$1.1$ & $3.47$ & $2.75$ & $-1.5$ & $0.472$ & $0.533$ & $0.303$ & $1.18$ & $1.23$ & $0.322$ & $1.08$ & $1.76$ & $1.21$ & $1.88$ \\
$1.2$ & $3.56$ & $3.08$ & $-1.77$ & $0.496$ & $0.556$ & $0.316$ & $1.2$ & $1.33$ & $0.358$ & $1.27$ & $1.76$ & $1.22$ & $2.03$ \\
	\hline 
\end{tabularx}
}	
	\caption{Numerical values of the various 4-loops NGLs integrals.} \label{tab:4loop_NumericalVals}	
\end{table}

\bibliographystyle{JHEP}
\bibliography{Refs}
\end{document}

%% file: Figs/G2.tex
\begingroup
  \makeatletter
  \providecommand\color[2][]{%
    \GenericError{(gnuplot) \space\space\space\@spaces}{%
      Package color not loaded in conjunction with
      terminal option `colourtext'%
    }{See the gnuplot documentation for explanation.%
    }{Either use 'blacktext' in gnuplot or load the package
      color.sty in LaTeX.}%
    \renewcommand\color[2][]{}%
  }%
  \providecommand\includegraphics[2][]{%
    \GenericError{(gnuplot) \space\space\space\@spaces}{%
      Package graphicx or graphics not loaded%
    }{See the gnuplot documentation for explanation.%
    }{The gnuplot epslatex terminal needs graphicx.sty or graphics.sty.}%
    \renewcommand\includegraphics[2][]{}%
  }%
  \providecommand\rotatebox[2]{#2}%
  \@ifundefined{ifGPcolor}{%
    \newif\ifGPcolor
    \GPcolortrue
  }{}%
  \@ifundefined{ifGPblacktext}{%
    \newif\ifGPblacktext
    \GPblacktexttrue
  }{}%
  \let\gplgaddtomacro\g@addto@macro
  \gdef\gplbacktext{}%
  \gdef\gplfronttext{}%
  \makeatother
  \ifGPblacktext
    \def\colorrgb#1{}%
    \def\colorgray#1{}%
  \else
    \ifGPcolor
      \def\colorrgb#1{\color[rgb]{#1}}%
      \def\colorgray#1{\color[gray]{#1}}%
      \expandafter\def\csname LTw\endcsname{\color{white}}%
      \expandafter\def\csname LTb\endcsname{\color{black}}%
      \expandafter\def\csname LTa\endcsname{\color{black}}%
      \expandafter\def\csname LT0\endcsname{\color[rgb]{1,0,0}}%
      \expandafter\def\csname LT1\endcsname{\color[rgb]{0,1,0}}%
      \expandafter\def\csname LT2\endcsname{\color[rgb]{0,0,1}}%
      \expandafter\def\csname LT3\endcsname{\color[rgb]{1,0,1}}%
      \expandafter\def\csname LT4\endcsname{\color[rgb]{0,1,1}}%
      \expandafter\def\csname LT5\endcsname{\color[rgb]{1,1,0}}%
      \expandafter\def\csname LT6\endcsname{\color[rgb]{0,0,0}}%
      \expandafter\def\csname LT7\endcsname{\color[rgb]{1,0.3,0}}%
      \expandafter\def\csname LT8\endcsname{\color[rgb]{0.5,0.5,0.5}}%
    \else
      \def\colorrgb#1{\color{black}}%
      \def\colorgray#1{\color[gray]{#1}}%
      \expandafter\def\csname LTw\endcsname{\color{white}}%
      \expandafter\def\csname LTb\endcsname{\color{black}}%
      \expandafter\def\csname LTa\endcsname{\color{black}}%
      \expandafter\def\csname LT0\endcsname{\color{black}}%
      \expandafter\def\csname LT1\endcsname{\color{black}}%
      \expandafter\def\csname LT2\endcsname{\color{black}}%
      \expandafter\def\csname LT3\endcsname{\color{black}}%
      \expandafter\def\csname LT4\endcsname{\color{black}}%
      \expandafter\def\csname LT5\endcsname{\color{black}}%
      \expandafter\def\csname LT6\endcsname{\color{black}}%
      \expandafter\def\csname LT7\endcsname{\color{black}}%
      \expandafter\def\csname LT8\endcsname{\color{black}}%
    \fi
  \fi
    \setlength{\unitlength}{0.0500bp}%
    \ifx\gptboxheight\undefined%
      \newlength{\gptboxheight}%
      \newlength{\gptboxwidth}%
      \newsavebox{\gptboxtext}%
    \fi%
    \setlength{\fboxrule}{0.5pt}%
    \setlength{\fboxsep}{1pt}%
\begin{picture}(7200.00,5040.00)%
    \gplgaddtomacro\gplbacktext{%
      \csname LTb\endcsname
      \put(946,704){\makebox(0,0)[r]{\strut{}\small 0}}%
      \put(946,1439){\makebox(0,0)[r]{\strut{}\small 5}}%
      \put(946,2174){\makebox(0,0)[r]{\strut{}\small 10}}%
      \put(946,2909){\makebox(0,0)[r]{\strut{}\small 15}}%
      \put(946,3644){\makebox(0,0)[r]{\strut{}\small 20}}%
      \put(946,4379){\makebox(0,0)[r]{\strut{}\small 25}}%
      \put(1078,484){\makebox(0,0){\strut{}\small 0}}%
      \put(1797,484){\makebox(0,0){\strut{}\small 0.2}}%
      \put(2517,484){\makebox(0,0){\strut{}\small 0.4}}%
      \put(3236,484){\makebox(0,0){\strut{}\small 0.6}}%
      \put(3955,484){\makebox(0,0){\strut{}\small 0.8}}%
      \put(4675,484){\makebox(0,0){\strut{}\small 1}}%
      \put(5394,484){\makebox(0,0){\strut{}\small 1.2}}%
      \put(6113,484){\makebox(0,0){\strut{}\small 1.4}}%
    }%
    \gplgaddtomacro\gplfronttext{%
      \csname LTb\endcsname
      \put(209,2541){\rotatebox{-270}{\makebox(0,0){\strut{}$\mathcal{G}_{2, \delta}(R)$}}}%
      \put(6638,2541){\rotatebox{-270}{\makebox(0,0){\strut{} }}}%
      \put(3775,154){\makebox(0,0){\strut{}$R$}}%
      \put(3775,4709){\makebox(0,0){\strut{} }}%
      \csname LTb\endcsname
      \put(2065,4206){\makebox(0,0)[l]{\strut{}($\delta_1$)}}%
      \csname LTb\endcsname
      \put(2065,3986){\makebox(0,0)[l]{\strut{}($\delta_2$)}}%
      \csname LTb\endcsname
      \put(2065,3766){\makebox(0,0)[l]{\strut{}($\delta_3$)}}%
    }%
    \gplbacktext
    \put(0,0){\includegraphics{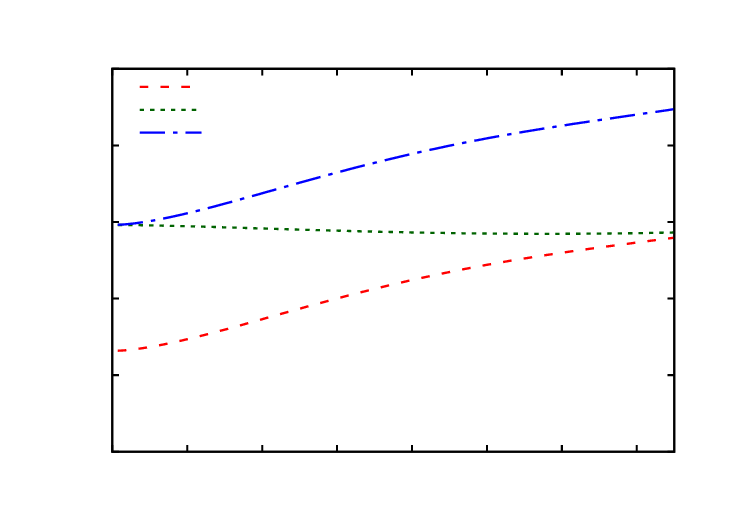}}%
    \gplfronttext
  \end{picture}%
\endgroup

%% file: Figs/G3.tex
\begingroup
  \makeatletter
  \providecommand\color[2][]{%
    \GenericError{(gnuplot) \space\space\space\@spaces}{%
      Package color not loaded in conjunction with
      terminal option `colourtext'%
    }{See the gnuplot documentation for explanation.%
    }{Either use 'blacktext' in gnuplot or load the package
      color.sty in LaTeX.}%
    \renewcommand\color[2][]{}%
  }%
  \providecommand\includegraphics[2][]{%
    \GenericError{(gnuplot) \space\space\space\@spaces}{%
      Package graphicx or graphics not loaded%
    }{See the gnuplot documentation for explanation.%
    }{The gnuplot epslatex terminal needs graphicx.sty or graphics.sty.}%
    \renewcommand\includegraphics[2][]{}%
  }%
  \providecommand\rotatebox[2]{#2}%
  \@ifundefined{ifGPcolor}{%
    \newif\ifGPcolor
    \GPcolortrue
  }{}%
  \@ifundefined{ifGPblacktext}{%
    \newif\ifGPblacktext
    \GPblacktexttrue
  }{}%
  \let\gplgaddtomacro\g@addto@macro
  \gdef\gplbacktext{}%
  \gdef\gplfronttext{}%
  \makeatother
  \ifGPblacktext
    \def\colorrgb#1{}%
    \def\colorgray#1{}%
  \else
    \ifGPcolor
      \def\colorrgb#1{\color[rgb]{#1}}%
      \def\colorgray#1{\color[gray]{#1}}%
      \expandafter\def\csname LTw\endcsname{\color{white}}%
      \expandafter\def\csname LTb\endcsname{\color{black}}%
      \expandafter\def\csname LTa\endcsname{\color{black}}%
      \expandafter\def\csname LT0\endcsname{\color[rgb]{1,0,0}}%
      \expandafter\def\csname LT1\endcsname{\color[rgb]{0,1,0}}%
      \expandafter\def\csname LT2\endcsname{\color[rgb]{0,0,1}}%
      \expandafter\def\csname LT3\endcsname{\color[rgb]{1,0,1}}%
      \expandafter\def\csname LT4\endcsname{\color[rgb]{0,1,1}}%
      \expandafter\def\csname LT5\endcsname{\color[rgb]{1,1,0}}%
      \expandafter\def\csname LT6\endcsname{\color[rgb]{0,0,0}}%
      \expandafter\def\csname LT7\endcsname{\color[rgb]{1,0.3,0}}%
      \expandafter\def\csname LT8\endcsname{\color[rgb]{0.5,0.5,0.5}}%
    \else
      \def\colorrgb#1{\color{black}}%
      \def\colorgray#1{\color[gray]{#1}}%
      \expandafter\def\csname LTw\endcsname{\color{white}}%
      \expandafter\def\csname LTb\endcsname{\color{black}}%
      \expandafter\def\csname LTa\endcsname{\color{black}}%
      \expandafter\def\csname LT0\endcsname{\color{black}}%
      \expandafter\def\csname LT1\endcsname{\color{black}}%
      \expandafter\def\csname LT2\endcsname{\color{black}}%
      \expandafter\def\csname LT3\endcsname{\color{black}}%
      \expandafter\def\csname LT4\endcsname{\color{black}}%
      \expandafter\def\csname LT5\endcsname{\color{black}}%
      \expandafter\def\csname LT6\endcsname{\color{black}}%
      \expandafter\def\csname LT7\endcsname{\color{black}}%
      \expandafter\def\csname LT8\endcsname{\color{black}}%
    \fi
  \fi
    \setlength{\unitlength}{0.0500bp}%
    \ifx\gptboxheight\undefined%
      \newlength{\gptboxheight}%
      \newlength{\gptboxwidth}%
      \newsavebox{\gptboxtext}%
    \fi%
    \setlength{\fboxrule}{0.5pt}%
    \setlength{\fboxsep}{1pt}%
\begin{picture}(7200.00,5040.00)%
    \gplgaddtomacro\gplbacktext{%
      \csname LTb\endcsname
      \put(946,704){\makebox(0,0)[r]{\strut{}\small 0}}%
      \put(946,1229){\makebox(0,0)[r]{\strut{}\small 10}}%
      \put(946,1754){\makebox(0,0)[r]{\strut{}\small 20}}%
      \put(946,2279){\makebox(0,0)[r]{\strut{}\small 30}}%
      \put(946,2804){\makebox(0,0)[r]{\strut{}\small 40}}%
      \put(946,3329){\makebox(0,0)[r]{\strut{}\small 50}}%
      \put(946,3854){\makebox(0,0)[r]{\strut{}\small 60}}%
      \put(946,4379){\makebox(0,0)[r]{\strut{}\small 70}}%
      \put(1078,484){\makebox(0,0){\strut{}\small 0}}%
      \put(1797,484){\makebox(0,0){\strut{}\small 0.2}}%
      \put(2517,484){\makebox(0,0){\strut{}\small 0.4}}%
      \put(3236,484){\makebox(0,0){\strut{}\small 0.6}}%
      \put(3955,484){\makebox(0,0){\strut{}\small 0.8}}%
      \put(4675,484){\makebox(0,0){\strut{}\small 1}}%
      \put(5394,484){\makebox(0,0){\strut{}\small 1.2}}%
      \put(6113,484){\makebox(0,0){\strut{}\small 1.4}}%
    }%
    \gplgaddtomacro\gplfronttext{%
      \csname LTb\endcsname
      \put(209,2541){\rotatebox{-270}{\makebox(0,0){\strut{}$\mathcal{G}_{3, \delta}(R)$}}}%
      \put(6638,2541){\rotatebox{-270}{\makebox(0,0){\strut{} }}}%
      \put(3775,154){\makebox(0,0){\strut{}$R$}}%
      \put(3775,4709){\makebox(0,0){\strut{} }}%
      \csname LTb\endcsname
      \put(2065,4206){\makebox(0,0)[l]{\strut{}($\delta_1$)}}%
      \csname LTb\endcsname
      \put(2065,3986){\makebox(0,0)[l]{\strut{}($\delta_2$)}}%
      \csname LTb\endcsname
      \put(2065,3766){\makebox(0,0)[l]{\strut{}($\delta_3$)}}%
    }%
    \gplbacktext
    \put(0,0){\includegraphics{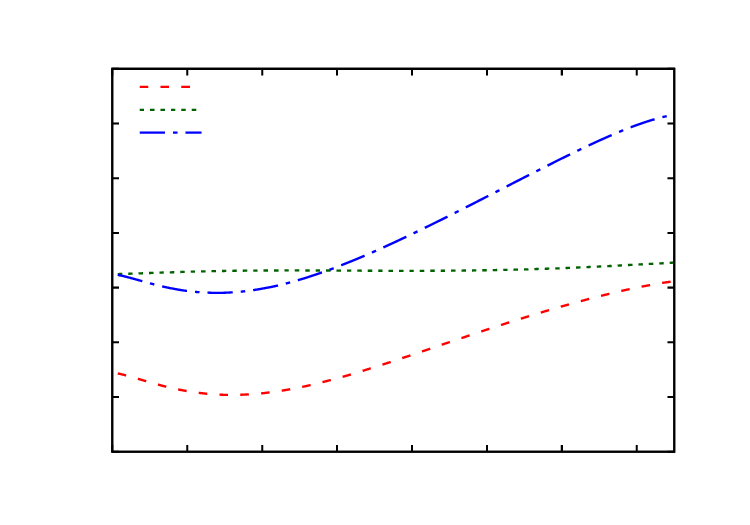}}%
    \gplfronttext
  \end{picture}%
\endgroup

%% file: Figs/G4l.tex
\begingroup
  \makeatletter
  \providecommand\color[2][]{%
    \GenericError{(gnuplot) \space\space\space\@spaces}{%
      Package color not loaded in conjunction with
      terminal option `colourtext'%
    }{See the gnuplot documentation for explanation.%
    }{Either use 'blacktext' in gnuplot or load the package
      color.sty in LaTeX.}%
    \renewcommand\color[2][]{}%
  }%
  \providecommand\includegraphics[2][]{%
    \GenericError{(gnuplot) \space\space\space\@spaces}{%
      Package graphicx or graphics not loaded%
    }{See the gnuplot documentation for explanation.%
    }{The gnuplot epslatex terminal needs graphicx.sty or graphics.sty.}%
    \renewcommand\includegraphics[2][]{}%
  }%
  \providecommand\rotatebox[2]{#2}%
  \@ifundefined{ifGPcolor}{%
    \newif\ifGPcolor
    \GPcolortrue
  }{}%
  \@ifundefined{ifGPblacktext}{%
    \newif\ifGPblacktext
    \GPblacktexttrue
  }{}%
  \let\gplgaddtomacro\g@addto@macro
  \gdef\gplbacktext{}%
  \gdef\gplfronttext{}%
  \makeatother
  \ifGPblacktext
    \def\colorrgb#1{}%
    \def\colorgray#1{}%
  \else
    \ifGPcolor
      \def\colorrgb#1{\color[rgb]{#1}}%
      \def\colorgray#1{\color[gray]{#1}}%
      \expandafter\def\csname LTw\endcsname{\color{white}}%
      \expandafter\def\csname LTb\endcsname{\color{black}}%
      \expandafter\def\csname LTa\endcsname{\color{black}}%
      \expandafter\def\csname LT0\endcsname{\color[rgb]{1,0,0}}%
      \expandafter\def\csname LT1\endcsname{\color[rgb]{0,1,0}}%
      \expandafter\def\csname LT2\endcsname{\color[rgb]{0,0,1}}%
      \expandafter\def\csname LT3\endcsname{\color[rgb]{1,0,1}}%
      \expandafter\def\csname LT4\endcsname{\color[rgb]{0,1,1}}%
      \expandafter\def\csname LT5\endcsname{\color[rgb]{1,1,0}}%
      \expandafter\def\csname LT6\endcsname{\color[rgb]{0,0,0}}%
      \expandafter\def\csname LT7\endcsname{\color[rgb]{1,0.3,0}}%
      \expandafter\def\csname LT8\endcsname{\color[rgb]{0.5,0.5,0.5}}%
    \else
      \def\colorrgb#1{\color{black}}%
      \def\colorgray#1{\color[gray]{#1}}%
      \expandafter\def\csname LTw\endcsname{\color{white}}%
      \expandafter\def\csname LTb\endcsname{\color{black}}%
      \expandafter\def\csname LTa\endcsname{\color{black}}%
      \expandafter\def\csname LT0\endcsname{\color{black}}%
      \expandafter\def\csname LT1\endcsname{\color{black}}%
      \expandafter\def\csname LT2\endcsname{\color{black}}%
      \expandafter\def\csname LT3\endcsname{\color{black}}%
      \expandafter\def\csname LT4\endcsname{\color{black}}%
      \expandafter\def\csname LT5\endcsname{\color{black}}%
      \expandafter\def\csname LT6\endcsname{\color{black}}%
      \expandafter\def\csname LT7\endcsname{\color{black}}%
      \expandafter\def\csname LT8\endcsname{\color{black}}%
    \fi
  \fi
    \setlength{\unitlength}{0.0500bp}%
    \ifx\gptboxheight\undefined%
      \newlength{\gptboxheight}%
      \newlength{\gptboxwidth}%
      \newsavebox{\gptboxtext}%
    \fi%
    \setlength{\fboxrule}{0.5pt}%
    \setlength{\fboxsep}{1pt}%
\begin{picture}(7200.00,5040.00)%
    \gplgaddtomacro\gplbacktext{%
      \csname LTb\endcsname
      \put(1078,704){\makebox(0,0)[r]{\strut{}\small 0}}%
      \put(1078,1372){\makebox(0,0)[r]{\strut{}\small 100}}%
      \put(1078,2040){\makebox(0,0)[r]{\strut{}\small 200}}%
      \put(1078,2709){\makebox(0,0)[r]{\strut{}\small 300}}%
      \put(1078,3377){\makebox(0,0)[r]{\strut{}\small 400}}%
      \put(1078,4045){\makebox(0,0)[r]{\strut{}\small 500}}%
      \put(1210,484){\makebox(0,0){\strut{}\small 0}}%
      \put(1912,484){\makebox(0,0){\strut{}\small 0.2}}%
      \put(2613,484){\makebox(0,0){\strut{}\small 0.4}}%
      \put(3315,484){\makebox(0,0){\strut{}\small 0.6}}%
      \put(4017,484){\makebox(0,0){\strut{}\small 0.8}}%
      \put(4719,484){\makebox(0,0){\strut{}\small 1}}%
      \put(5420,484){\makebox(0,0){\strut{}\small 1.2}}%
      \put(6122,484){\makebox(0,0){\strut{}\small 1.4}}%
    }%
    \gplgaddtomacro\gplfronttext{%
      \csname LTb\endcsname
      \put(209,2541){\rotatebox{-270}{\makebox(0,0){\strut{}$\mathcal{G}_{4, \delta}(R)$}}}%
      \put(6638,2541){\rotatebox{-270}{\makebox(0,0){\strut{} }}}%
      \put(3841,154){\makebox(0,0){\strut{}$R$}}%
      \put(3841,4709){\makebox(0,0){\strut{} }}%
      \csname LTb\endcsname
      \put(2197,4206){\makebox(0,0)[l]{\strut{}($\delta_1$)}}%
      \csname LTb\endcsname
      \put(2197,3986){\makebox(0,0)[l]{\strut{}($\delta_2$)}}%
      \csname LTb\endcsname
      \put(2197,3766){\makebox(0,0)[l]{\strut{}($\delta_3$)}}%
    }%
    \gplbacktext
    \put(0,0){\includegraphics{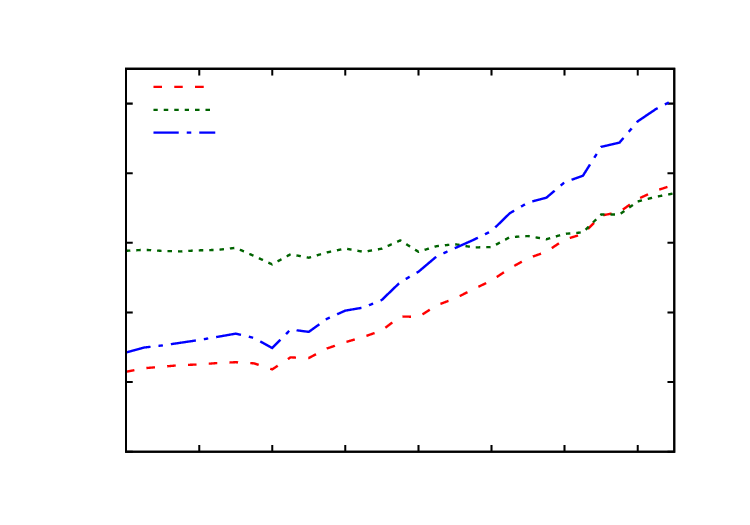}}%
    \gplfronttext
  \end{picture}%
\endgroup

%% file: VHjet_akt.bbl
\providecommand{\href}[2]{#2}\begingroup\raggedright\begin{thebibliography}{10}

\bibitem{Khelifa-Kerfa:2015mma}
K.~Khelifa-Kerfa and Y.~Delenda, \emph{{Non-global logarithms at finite N$_{c}$
  beyond leading order}},
  \href{https://doi.org/10.1007/JHEP03(2015)094}{\emph{JHEP} {\bfseries 03}
  (2015) 094} [\href{https://arxiv.org/abs/1501.00475}{{\ttfamily
  1501.00475}}].

\bibitem{Kogler:2018hem}
R.~Kogler et~al., \emph{{Jet Substructure at the Large Hadron Collider:
  Experimental Review}},
  \href{https://doi.org/10.1103/RevModPhys.91.045003}{\emph{Rev. Mod. Phys.}
  {\bfseries 91} (2019) 045003}
  [\href{https://arxiv.org/abs/1803.06991}{{\ttfamily 1803.06991}}].

\bibitem{Larkoski:2017jix}
A.J.~Larkoski, I.~Moult and B.~Nachman, \emph{{Jet Substructure at the Large
  Hadron Collider: A Review of Recent Advances in Theory and Machine
  Learning}}, \href{https://doi.org/10.1016/j.physrep.2019.11.001}{\emph{Phys.
  Rept.} {\bfseries 841} (2020) 1}
  [\href{https://arxiv.org/abs/1709.04464}{{\ttfamily 1709.04464}}].

\bibitem{Dasgupta:2001sh}
M.~Dasgupta and G.P.~Salam, \emph{{Resummation of nonglobal QCD observables}},
  \href{https://doi.org/10.1016/S0370-2693(01)00725-0}{\emph{Phys. Lett. B}
  {\bfseries 512} (2001) 323}
  [\href{https://arxiv.org/abs/hep-ph/0104277}{{\ttfamily hep-ph/0104277}}].

\bibitem{Dasgupta:2002bw}
M.~Dasgupta and G.P.~Salam, \emph{{Accounting for coherence in interjet E(t)
  flow: A Case study}},
  \href{https://doi.org/10.1088/1126-6708/2002/03/017}{\emph{JHEP} {\bfseries
  03} (2002) 017} [\href{https://arxiv.org/abs/hep-ph/0203009}{{\ttfamily
  hep-ph/0203009}}].

\bibitem{Dasgupta:2002dc}
M.~Dasgupta and G.P.~Salam, \emph{{Resummed event shape variables in DIS}},
  \href{https://doi.org/10.1088/1126-6708/2002/08/032}{\emph{JHEP} {\bfseries
  08} (2002) 032} [\href{https://arxiv.org/abs/hep-ph/0208073}{{\ttfamily
  hep-ph/0208073}}].

\bibitem{Banfi:2002hw}
A.~Banfi, G.~Marchesini and G.~Smye, \emph{{Away from jet energy flow}},
  \href{https://doi.org/10.1088/1126-6708/2002/08/006}{\emph{JHEP} {\bfseries
  08} (2002) 006} [\href{https://arxiv.org/abs/hep-ph/0206076}{{\ttfamily
  hep-ph/0206076}}].

\bibitem{Appleby:2002ke}
R.B.~Appleby and M.H.~Seymour, \emph{{Nonglobal logarithms in interjet energy
  flow with k$_t$ clustering requirement}},
  \href{https://doi.org/10.1088/1126-6708/2002/12/063}{\emph{JHEP} {\bfseries
  12} (2002) 063} [\href{https://arxiv.org/abs/hep-ph/0211426}{{\ttfamily
  hep-ph/0211426}}].

\bibitem{Forshaw:2006fk}
J.R.~Forshaw, A.~Kyrieleis and M.H.~Seymour, \emph{{Super-leading logarithms in
  non-global observables in QCD}},
  \href{https://doi.org/10.1088/1126-6708/2006/08/059}{\emph{JHEP} {\bfseries
  08} (2006) 059} [\href{https://arxiv.org/abs/hep-ph/0604094}{{\ttfamily
  hep-ph/0604094}}].

\bibitem{Banfi:2010pa}
A.~Banfi, M.~Dasgupta, K.~Khelifa-Kerfa and S.~Marzani, \emph{{Non-global
  logarithms and jet algorithms in high-pT jet shapes}},
  \href{https://doi.org/10.1007/JHEP08(2010)064}{\emph{JHEP} {\bfseries 08}
  (2010) 064} [\href{https://arxiv.org/abs/1004.3483}{{\ttfamily 1004.3483}}].

\bibitem{Khelifa-Kerfa:2011quw}
K.~Khelifa-Kerfa, \emph{{Non-global logs and clustering impact on jet mass with
  a jet veto distribution}},
  \href{https://doi.org/10.1007/JHEP02(2012)072}{\emph{JHEP} {\bfseries 02}
  (2012) 072} [\href{https://arxiv.org/abs/1111.2016}{{\ttfamily 1111.2016}}].

\bibitem{Dasgupta:2012hg}
M.~Dasgupta, K.~Khelifa-Kerfa, S.~Marzani and M.~Spannowsky, \emph{{On jet mass
  distributions in Z+jet and dijet processes at the LHC}},
  \href{https://doi.org/10.1007/JHEP10(2012)126}{\emph{JHEP} {\bfseries 10}
  (2012) 126} [\href{https://arxiv.org/abs/1207.1640}{{\ttfamily 1207.1640}}].

\bibitem{Delenda:2012mm}
Y.~Delenda and K.~Khelifa-Kerfa, \emph{{On the resummation of clustering
  logarithms for non-global observables}},
  \href{https://doi.org/10.1007/JHEP09(2012)109}{\emph{JHEP} {\bfseries 09}
  (2012) 109} [\href{https://arxiv.org/abs/1207.4528}{{\ttfamily 1207.4528}}].

\bibitem{Schwartz:2014wha}
M.D.~Schwartz and H.X.~Zhu, \emph{{Nonglobal logarithms at three loops, four
  loops, five loops, and beyond}},
  \href{https://doi.org/10.1103/PhysRevD.90.065004}{\emph{Phys. Rev. D}
  {\bfseries 90} (2014) 065004}
  [\href{https://arxiv.org/abs/1403.4949}{{\ttfamily 1403.4949}}].

\bibitem{Banfi:2021owj}
A.~Banfi, F.A.~Dreyer and P.F.~Monni, \emph{{Next-to-leading non-global
  logarithms in QCD}},
  \href{https://doi.org/10.1007/JHEP10(2021)006}{\emph{JHEP} {\bfseries 10}
  (2021) 006} [\href{https://arxiv.org/abs/2104.06416}{{\ttfamily
  2104.06416}}].

\bibitem{Banfi:2021xzn}
A.~Banfi, F.A.~Dreyer and P.F.~Monni, \emph{{Higher-order non-global logarithms
  from jet calculus}},
  \href{https://doi.org/10.1007/JHEP03(2022)135}{\emph{JHEP} {\bfseries 03}
  (2022) 135} [\href{https://arxiv.org/abs/2111.02413}{{\ttfamily
  2111.02413}}].

\bibitem{Becher:2021urs}
T.~Becher, T.~Rauh and X.~Xu, \emph{{Two-loop anomalous dimension for the
  resummation of non-global observables}},
  \href{https://doi.org/10.1007/JHEP08(2022)134}{\emph{JHEP} {\bfseries 08}
  (2022) 134} [\href{https://arxiv.org/abs/2112.02108}{{\ttfamily
  2112.02108}}].

\bibitem{Becher:2023vrh}
T.~Becher, N.~Schalch and X.~Xu, \emph{{Resummation of Next-to-Leading
  Nonglobal Logarithms at the LHC}},
  \href{https://doi.org/10.1103/PhysRevLett.132.081602}{\emph{Phys. Rev. Lett.}
  {\bfseries 132} (2024) 081602}
  [\href{https://arxiv.org/abs/2307.02283}{{\ttfamily 2307.02283}}].

\bibitem{vanBeekveld:2023lsa}
M.~van Beekveld, M.~Dasgupta, B.K.~El-Menoufi, J.~Helliwell and P.F.~Monni,
  \emph{{Collinear fragmentation at NNLL: generating functionals, groomed
  correlators and angularities}},
  \href{https://doi.org/10.1007/JHEP05(2024)093}{\emph{JHEP} {\bfseries 05}
  (2024) 093} [\href{https://arxiv.org/abs/2307.15734}{{\ttfamily
  2307.15734}}].

\bibitem{vanBeekveld:2023ivn}
M.~van Beekveld et~al., \emph{{Introduction to the PanScales framework, version
  0.1}},  \href{https://arxiv.org/abs/2312.13275}{{\ttfamily 2312.13275}}.

\bibitem{FerrarioRavasio:2023kyg}
S.~Ferrario~Ravasio, K.~Hamilton, A.~Karlberg, G.P.~Salam, L.~Scyboz and
  G.~Soyez, \emph{{Parton Showering with Higher Logarithmic Accuracy for Soft
  Emissions}},
  \href{https://doi.org/10.1103/PhysRevLett.131.161906}{\emph{Phys. Rev. Lett.}
  {\bfseries 131} (2023) 161906}
  [\href{https://arxiv.org/abs/2307.11142}{{\ttfamily 2307.11142}}].

\bibitem{vanBeekveld:2024wws}
M.~van Beekveld et~al., \emph{{A new standard for the logarithmic accuracy of
  parton showers}},  \href{https://arxiv.org/abs/2406.02661}{{\ttfamily
  2406.02661}}.

\bibitem{Hatta:2013iba}
Y.~Hatta and T.~Ueda, \emph{{Resummation of non-global logarithms at finite
  $N_c$}}, \href{https://doi.org/10.1016/j.nuclphysb.2013.06.021}{\emph{Nucl.
  Phys. B} {\bfseries 874} (2013) 808}
  [\href{https://arxiv.org/abs/1304.6930}{{\ttfamily 1304.6930}}].

\bibitem{Hatta:2020wre}
Y.~Hatta and T.~Ueda, \emph{{Non-global logarithms in hadron collisions at
  $N_c$ = 3}},
  \href{https://doi.org/10.1016/j.nuclphysb.2020.115273}{\emph{Nucl. Phys. B}
  {\bfseries 962} (2021) 115273}
  [\href{https://arxiv.org/abs/2011.04154}{{\ttfamily 2011.04154}}].

\bibitem{Hagiwara:2015bia}
Y.~Hagiwara, Y.~Hatta and T.~Ueda, \emph{{Hemisphere jet mass distribution at
  finite $N_c$}},
  \href{https://doi.org/10.1016/j.physletb.2016.03.028}{\emph{Phys. Lett. B}
  {\bfseries 756} (2016) 254}
  [\href{https://arxiv.org/abs/1507.07641}{{\ttfamily 1507.07641}}].

\bibitem{Platzer:2013fha}
S.~Pl\"atzer, \emph{{Summing Large-$N$ Towers in Colour Flow Evolution}},
  \href{https://doi.org/10.1140/epjc/s10052-014-2907-2}{\emph{Eur. Phys. J. C}
  {\bfseries 74} (2014) 2907}
  [\href{https://arxiv.org/abs/1312.2448}{{\ttfamily 1312.2448}}].

\bibitem{Forshaw:2019ver}
J.R.~Forshaw, J.~Holguin and S.~Pl\"atzer, \emph{{Parton branching at amplitude
  level}}, \href{https://doi.org/10.1007/JHEP08(2019)145}{\emph{JHEP}
  {\bfseries 08} (2019) 145}
  [\href{https://arxiv.org/abs/1905.08686}{{\ttfamily 1905.08686}}].

\bibitem{AngelesMartinez:2018cfz}
R.~\'Angeles~Mart\'\i{}nez, M.~De~Angelis, J.R.~Forshaw, S.~Pl\"atzer and
  M.H.~Seymour, \emph{{Soft gluon evolution and non-global logarithms}},
  \href{https://doi.org/10.1007/JHEP05(2018)044}{\emph{JHEP} {\bfseries 05}
  (2018) 044} [\href{https://arxiv.org/abs/1802.08531}{{\ttfamily
  1802.08531}}].

\bibitem{DeAngelis:2020rvq}
M.~De~Angelis, J.R.~Forshaw and S.~Pl\"atzer, \emph{{Resummation and Simulation
  of Soft Gluon Effects beyond Leading Color}},
  \href{https://doi.org/10.1103/PhysRevLett.126.112001}{\emph{Phys. Rev. Lett.}
  {\bfseries 126} (2021) 112001}
  [\href{https://arxiv.org/abs/2007.09648}{{\ttfamily 2007.09648}}].

\bibitem{Becher:2023znt}
T.~Becher and J.~Haag, \emph{{Factorization and resummation for sequential
  recombination jet cross sections}},
  \href{https://doi.org/10.1007/JHEP01(2024)155}{\emph{JHEP} {\bfseries 01}
  (2024) 155} [\href{https://arxiv.org/abs/2309.17355}{{\ttfamily
  2309.17355}}].

\bibitem{Cacciari:2008gp}
M.~Cacciari, G.P.~Salam and G.~Soyez, \emph{{The anti-k$_t$ jet clustering
  algorithm}}, \href{https://doi.org/10.1088/1126-6708/2008/04/063}{\emph{JHEP}
  {\bfseries 04} (2008) 063} [\href{https://arxiv.org/abs/0802.1189}{{\ttfamily
  0802.1189}}].

\bibitem{Catani:1993hr}
S.~Catani, Y.L.~Dokshitzer, M.H.~Seymour and B.R.~Webber, \emph{{Longitudinally
  invariant $K_t$ clustering algorithms for hadron hadron collisions}},
  \href{https://doi.org/10.1016/0550-3213(93)90166-M}{\emph{Nucl. Phys. B}
  {\bfseries 406} (1993) 187}.

\bibitem{Ellis:1993tq}
S.D.~Ellis and D.E.~Soper, \emph{{Successive combination jet algorithm for
  hadron collisions}},
  \href{https://doi.org/10.1103/PhysRevD.48.3160}{\emph{Phys. Rev.} {\bfseries
  D48} (1993) 3160} [\href{https://arxiv.org/abs/hep-ph/9305266}{{\ttfamily
  hep-ph/9305266}}].

\bibitem{Dokshitzer:1997in}
Y.L.~Dokshitzer, G.D.~Leder, S.~Moretti and B.R.~Webber, \emph{{Better jet
  clustering algorithms}},
  \href{https://doi.org/10.1088/1126-6708/1997/08/001}{\emph{JHEP} {\bfseries
  08} (1997) 001} [\href{https://arxiv.org/abs/hep-ph/9707323}{{\ttfamily
  hep-ph/9707323}}].

\bibitem{Ziani:2021dxr}
N.~Ziani, K.~Khelifa-Kerfa and Y.~Delenda, \emph{{Jet mass distribution in
  Higgs/vector boson + jet events at hadron colliders with $k_t$ clustering}},
  \href{https://doi.org/10.1140/epjc/s10052-021-09379-z}{\emph{Eur. Phys. J. C}
  {\bfseries 81} (2021) 570}
  [\href{https://arxiv.org/abs/2104.11060}{{\ttfamily 2104.11060}}].

\bibitem{Gauld:2021ule}
R.~Gauld, A.~Gehrmann-De~Ridder, E.W.N.~Glover, A.~Huss and I.~Majer, \emph{{VH
  + jet production in hadron-hadron collisions up to order $
  {\alpha}_{\mathrm{s}}^3 $ in perturbative QCD}},
  \href{https://doi.org/10.1007/JHEP03(2022)008}{\emph{JHEP} {\bfseries 03}
  (2022) 008} [\href{https://arxiv.org/abs/2110.12992}{{\ttfamily
  2110.12992}}].

\bibitem{Boughezal:2015aha}
R.~Boughezal, C.~Focke, W.~Giele, X.~Liu and F.~Petriello, \emph{{Higgs boson
  production in association with a jet at NNLO using jettiness subtraction}},
  \href{https://doi.org/10.1016/j.physletb.2015.06.055}{\emph{Phys. Lett. B}
  {\bfseries 748} (2015) 5} [\href{https://arxiv.org/abs/1505.03893}{{\ttfamily
  1505.03893}}].

\bibitem{Boughezal:2015ded}
R.~Boughezal, J.M.~Campbell, R.K.~Ellis, C.~Focke, W.T.~Giele, X.~Liu et~al.,
  \emph{{Z-boson production in association with a jet at
  next-to-next-to-leading order in perturbative QCD}},
  \href{https://doi.org/10.1103/PhysRevLett.116.152001}{\emph{Phys. Rev. Lett.}
  {\bfseries 116} (2016) 152001}
  [\href{https://arxiv.org/abs/1512.01291}{{\ttfamily 1512.01291}}].

\bibitem{Gehrmann-DeRidder:2016cdi}
A.~Gehrmann-De~Ridder, T.~Gehrmann, E.W.N.~Glover, A.~Huss and T.A.~Morgan,
  \emph{{The NNLO QCD corrections to Z boson production at large transverse
  momentum}}, \href{https://doi.org/10.1007/JHEP07(2016)133}{\emph{JHEP}
  {\bfseries 07} (2016) 133}
  [\href{https://arxiv.org/abs/1605.04295}{{\ttfamily 1605.04295}}].

\bibitem{Boughezal:2016dtm}
R.~Boughezal, X.~Liu and F.~Petriello, \emph{{W-boson plus jet differential
  distributions at NNLO in QCD}},
  \href{https://doi.org/10.1103/PhysRevD.94.113009}{\emph{Phys. Rev. D}
  {\bfseries 94} (2016) 113009}
  [\href{https://arxiv.org/abs/1602.06965}{{\ttfamily 1602.06965}}].

\bibitem{Gehrmann-DeRidder:2015wbt}
A.~Gehrmann-De~Ridder, T.~Gehrmann, E.W.N.~Glover, A.~Huss and T.A.~Morgan,
  \emph{{Precise QCD predictions for the production of a Z boson in association
  with a hadronic jet}},
  \href{https://doi.org/10.1103/PhysRevLett.117.022001}{\emph{Phys. Rev. Lett.}
  {\bfseries 117} (2016) 022001}
  [\href{https://arxiv.org/abs/1507.02850}{{\ttfamily 1507.02850}}].

\bibitem{Campbell:2017dqk}
J.M.~Campbell, R.K.~Ellis and C.~Williams, \emph{{Driving missing data at the
  LHC: NNLO predictions for the ratio of $\gamma+j$ and $Z+j$}},
  \href{https://doi.org/10.1103/PhysRevD.96.014037}{\emph{Phys. Rev. D}
  {\bfseries 96} (2017) 014037}
  [\href{https://arxiv.org/abs/1703.10109}{{\ttfamily 1703.10109}}].

\bibitem{Khelifa-Kerfa:2020nlc}
K.~Khelifa-Kerfa and Y.~Delenda, \emph{{Eikonal amplitudes for three-hard legs
  processes at finite-N$_c$}},
  \href{https://doi.org/10.1016/j.physletb.2020.135768}{\emph{Phys. Lett. B}
  {\bfseries 809} (2020) 135768}
  [\href{https://arxiv.org/abs/2006.08758}{{\ttfamily 2006.08758}}].

\bibitem{Delenda:2015tbo}
Y.~Delenda and K.~Khelifa-Kerfa, \emph{{Eikonal gluon bremsstrahlung at finite
  $N_c$ beyond two loops}},
  \href{https://doi.org/10.1103/PhysRevD.93.054027}{\emph{Phys. Rev. D}
  {\bfseries 93} (2016) 054027}
  [\href{https://arxiv.org/abs/1512.05401}{{\ttfamily 1512.05401}}].

\bibitem{Hahn:2004fe}
T.~Hahn, \emph{{CUBA: A Library for multidimensional numerical integration}},
  \href{https://doi.org/10.1016/j.cpc.2005.01.010}{\emph{Comput. Phys. Commun.}
  {\bfseries 168} (2005) 78}
  [\href{https://arxiv.org/abs/hep-ph/0404043}{{\ttfamily hep-ph/0404043}}].

\bibitem{Hahn:2016ktb}
T.~Hahn, \emph{{Concurrent Cuba}},
  \href{https://doi.org/10.1016/j.cpc.2016.05.012}{\emph{Comput. Phys. Commun.}
  {\bfseries 207} (2016) 341}.

\bibitem{VHjet_kt_4loop}
K.~Khelifa-Kerfa, ``{NGLs up to 4-loops for V/H+jet processes in k$_t$
  algorithm}.'' {in progress}.

\bibitem{Becher:2021zkk}
T.~Becher, M.~Neubert and D.Y.~Shao, \emph{{Resummation of Super-Leading
  Logarithms}},
  \href{https://doi.org/10.1103/PhysRevLett.127.212002}{\emph{Phys. Rev. Lett.}
  {\bfseries 127} (2021) 212002}
  [\href{https://arxiv.org/abs/2107.01212}{{\ttfamily 2107.01212}}].

\bibitem{Becher:2023mtx}
T.~Becher, M.~Neubert, D.Y.~Shao and M.~Stillger, \emph{{Factorization of
  non-global LHC observables and resummation of super-leading logarithms}},
  \href{https://doi.org/10.1007/JHEP12(2023)116}{\emph{JHEP} {\bfseries 12}
  (2023) 116} [\href{https://arxiv.org/abs/2307.06359}{{\ttfamily
  2307.06359}}].

\bibitem{Forshaw:2008cq}
J.R.~Forshaw, A.~Kyrieleis and M.H.~Seymour, \emph{{Super-leading logarithms in
  non-global observables in QCD: Colour basis independent calculation}},
  \href{https://doi.org/10.1088/1126-6708/2008/09/128}{\emph{JHEP} {\bfseries
  09} (2008) 128} [\href{https://arxiv.org/abs/0808.1269}{{\ttfamily
  0808.1269}}].

\bibitem{Keates:2009dn}
J.~Keates and M.H.~Seymour, \emph{{Super-leading logarithms in non-global
  observables in QCD: Fixed order calculation}},
  \href{https://doi.org/10.1088/1126-6708/2009/04/040}{\emph{JHEP} {\bfseries
  04} (2009) 040} [\href{https://arxiv.org/abs/0902.0477}{{\ttfamily
  0902.0477}}].

\bibitem{Banfi:2004yd}
A.~Banfi, G.P.~Salam and G.~Zanderighi, \emph{{Principles of general
  final-state resummation and automated implementation}},
  \href{https://doi.org/10.1088/1126-6708/2005/03/073}{\emph{JHEP} {\bfseries
  03} (2005) 073} [\href{https://arxiv.org/abs/hep-ph/0407286}{{\ttfamily
  hep-ph/0407286}}].

\bibitem{12loop}
S.~Caron-Huot, ``{Note on perturbative solution to the BMS equation}.''
  {private communication}.

\bibitem{Larkoski:2016zzc}
A.J.~Larkoski, I.~Moult and D.~Neill, \emph{{The Analytic Structure of
  Non-Global Logarithms: Convergence of the Dressed Gluon Expansion}},
  \href{https://doi.org/10.1007/JHEP11(2016)089}{\emph{JHEP} {\bfseries 11}
  (2016) 089} [\href{https://arxiv.org/abs/1609.04011}{{\ttfamily
  1609.04011}}].

\bibitem{Altarelli:1994vz}
G.~Altarelli, P.~Nason and G.~Ridolfi, \emph{{A Study of ultraviolet renormalon
  ambiguities in the determination of alpha-s from tau decay}},
  \href{https://doi.org/10.1007/BF01566673}{\emph{Z. Phys. C} {\bfseries 68}
  (1995) 257} [\href{https://arxiv.org/abs/hep-ph/9501240}{{\ttfamily
  hep-ph/9501240}}].

\bibitem{Caprini:2000js}
I.~Caprini and J.~Fischer, \emph{{Convergence of the expansion of the
  Laplace-Borel integral in perturbative QCD improved by conformal mapping}},
  \href{https://doi.org/10.1103/PhysRevD.62.054007}{\emph{Phys. Rev. D}
  {\bfseries 62} (2000) 054007}
  [\href{https://arxiv.org/abs/hep-ph/0002016}{{\ttfamily hep-ph/0002016}}].

\bibitem{Caprini:2010ir}
I.~Caprini and J.~Fischer, \emph{{Determination of $\alpha_s(M_\tau^2)$: a
  conformal mapping approach}},
  \href{https://doi.org/10.1016/j.nuclphysbps.2011.06.022}{\emph{Nucl. Phys. B
  Proc. Suppl.} {\bfseries 218} (2011) 128}
  [\href{https://arxiv.org/abs/1011.6480}{{\ttfamily 1011.6480}}].

\bibitem{Zinn-Justin:2002ecy}
J.~Zinn-Justin, \emph{{Quantum field theory and critical phenomena}},
  {\emph{Int. Ser. Monogr. Phys.} {\bfseries 113} (2002) 1}.

\end{thebibliography}\endgroup
